\documentclass[aip,jap,amsmath,amssymb,reprint]{revtex4-1}

\usepackage{bm,graphicx,epsfig,verbatim}
\usepackage{hyperref}

\begin{document}
\title{Photoinduced nonequilibrium dynamics in charge ordered materials}

\author{Linghua Zhu}
    \affiliation{Department of Physics, New Jersey Institute of Technology, Newark, New Jersey 07102 USA}
\author{Tsezar F. Seman}
    \affiliation{Department of Physics, Northern Illinois University, DeKalb, Illinois 60115 USA}
    \affiliation{Advanced Photon Source, Argonne National Laboratory, Argonne, Illinois 60439 USA}
\author{Michel van Veenendaal}
    \affiliation{Department of Physics, Northern Illinois University, DeKalb, Illinois 60115 USA}
    \affiliation{Advanced Photon Source, Argonne National Laboratory, Argonne, Illinois 60439 USA}
\author{Keun Hyuk Ahn}
    \email{kenahn@njit.edu}
    \affiliation{Department of Physics, New Jersey Institute of Technology, Newark, New Jersey 07102 USA}


\begin{abstract}
We study the nonequilibrium dynamics of photoinduced phase transitions in charge ordered (CO) systems with a strong electron-lattice interaction and analyze the interplay between electrons, periodic lattice distortions, and a phonon thermal reservoir. Simulations based on a tight-binding Hamiltonian and Boltzmann equations reveal partially decoupled oscillations of the electronic order parameter and the periodic lattice distortion during CO melting, which becomes more energy efficient with lower photon energy. The cooling rate of the electron system correlates with the CO gap dynamics, responsible for an order of magnitude decrease of the cooling rate upon the gap reopening. We also find that the time-dependent frequency of coherent oscillation reflects the dynamics of the energy landscape, such as transition between single-well and double-well, which sensitively depends on the photon energy and the pump fluence. The results demonstrate the intricate nonequilibrium dynamics in CO materials.
\end{abstract}


\maketitle

\section{Introduction}\label{sec:intro}

Advances in computing and communication technology demand ultrafast switching devices. Recently, photoinduced insulator-metal transitions in charge ordered (CO) or charge density wave (CDW) materials have been considered as a mechanism for future ultrafast switching devices.~\cite{Rohwer2011NATLET,Cilento2010JAP,Xue2013JAP} In addition, studies of photoinduced nonequilibrium dynamics have revealed properties and phases of materials inaccessible through equilibrium thermodynamic processes.~\cite{Zhang14REV,Kennes2017NAT} One class of materials of particular interest are transition metal oxides of perovskite or Ruddlesden-Popper structure, which include manganites, cuprates, and nickelates.~\cite{Matsubara2008JAP,Piazza2014STDYN,Mankowsky2015PRB,Caviglia2013PRB,Esposito2017PRL,Zhang2013JAP} The layers of $M$O$_{2}$, where $M$ and $\rm O$ represent a transition metal element and oxygen respectively, play a dominant role in electronic properties of these materials. For example, time-resolved experiments on $\rm Pr_{0.5}Ca_{0.5}MnO_{3}$ in a CO phase using ultrashort optical pump and x-ray probe at or off resonance have revealed decoupled nonequilibrium dynamics of electrons and periodic lattice distortion during photoinduced melting of the CO phase.~\cite{Beaud2014NATMAT}

In spite of the recent experimental progress, theoretical and computational studies of nonequilibrium dynamics in CO and related CDW materials have been restricted to phenomenological Ginzburg-Landau approaches,~\cite{Beaud2014NATMAT} calculations of carrier-doping effects using density functional theory,~\cite{Shao2016PRB} and models based on dynamics of the electronic density of states (DOS).~\cite{vanVeenendaal2013PRB}

In this paper, we present simulations of the photoinduced CO insulator-metal transitions in a model $M$$\rm O_{2}$ system, using a tight-binding Hamiltonian and a coupling between the electrons on $M$ ions and distortion of $\rm O$ ions. The dynamics of the periodic lattice distortion is treated classically. The electron dynamics follows the Boltzmann equations, as done in Refs.~\onlinecite{Groeneveld1992PRB,Groeneveld1995PRB,Rethfeld2002PRB,Demsar2003PRL,Ahn2004PRB,vanVeenendaal2013PRB}. The CO phase is recovered through the coupling between the electron system and a phonon thermal reservoir. Detailed time-domain studies of photoinduced melting of CO, particularly dynamics of the energy landscape, are presented. The results reveal nonequilibrium dynamics of the electronic order parameter and the periodic lattice distortion under various conditions of the photon energy and the pump fluence. In addition to the CO in transition metal oxides, the results are compared with experiments on CDW materials of other structures, because both phenomena involve coupled electron density modulation and lattice distortions.

The paper is organized as follows. Section~\ref{sec:model} presents the model system and equations governing the dynamics of the model system. Results of our simulations are shown in Sec.~\ref{sec:results} and compared with experimental results in Sec.~\ref{sec:comparison}. A summary is provided in Sec.~\ref{sec:summary}.

\section{Model}\label{sec:model}
\subsection{Hamiltonian}
\begin{figure}
    \centering
    \includegraphics[width=0.6\hsize,clip]{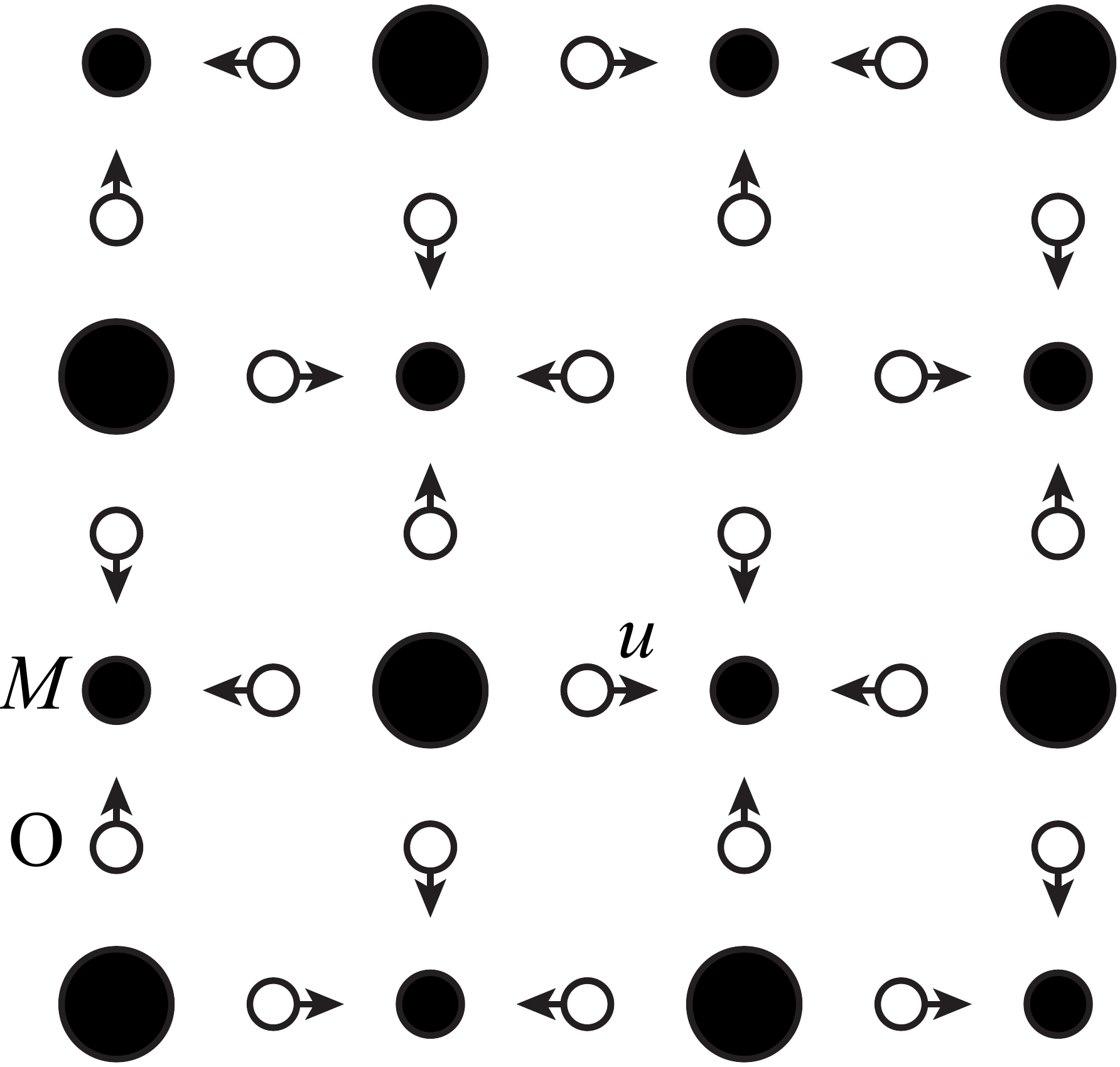}
    \caption{
    The model system of an $\it M\rm O_2$ square lattice with periodic boundary conditions. The size of the solid circles schematically represents the variation of electron density on $M$ ions in CO state. Arrows show the displacements of the $\rm O$ ions, represented by $u$.
    } \label{fig:model}
\end{figure}

We consider a model system of a $N{\times}N$ $M$${\rm O_{2}}$ square lattice with periodic boundary conditions, shown in Fig.~\ref{fig:model}. To capture the essential mechanism of CO transition in a model, we consider one spinless isotropic electron orbital per $M$ ion. The electron creation operator on the $M$ site at ${\bf n}=(n_x,n_y)$ is represented by $c^\dag_{\bf n}$. In this model, the $\rm CO$ instability arises as a result of Fermi surface nesting and electron-lattice coupling. Therefore, we include the displacements of the $\rm O$ ions at ${\bf n}+{\bf e}_{a}/2$ along the $a$-direction represented by $u^{a}_{\bf n}$ in the model, where ${a}=x,y$. One electron is present per two $M$ sites in the system, which would result in the checkerboard CO state and the lattice distortions shown in Fig.~\ref{fig:model}. The periodic distortion of the $\rm O$ ions is parameterized by a classical variable $u$, as indicated in Fig.~\ref{fig:model}. Motion of the $M$ ions is not considered because the O ions move symmetrically with respect to the $M$ ions.

The Hamiltonian for electrons has two terms. The first term represents the electron hopping between the nearest neighbor $M$ sites, given by
\begin{equation}
    H_{\rm hop} = -t_{h} \sum_{\bf n} \big( c_{\bf n}^\dag c_{{\bf n}+{\bf e_{\textit x}}} + c_{\bf n}^\dag c_{{\bf n}+{\bf e_{\textit y}}} + {\rm H.c.} \big),
\end{equation}
where $t_{h}$ is the electron hopping constant. The second term represents the coupling between the electron at $M$ site and the distortion of the surrounding negatively-charged $\rm O$ ions, given by
\begin{equation}
    H_{\rm el \mbox{-} latt} = -\lambda \sum_{\bf n} \frac{u_{\bf n}^x - u_{{\bf n}-{\bf e_{\textit x}}}^x + u_{\bf n}^y - u_{{\bf n}-{\bf e_{\textit y}}}^y}{4}  c_{\bf n}^\dag c_{\bf n} ,
\label{eq:Hel-la}
\end{equation}
where $\lambda$ is the electron-lattice coupling constant. The potential and the kinetic energies of the O ions are treated classically, and represented by the Hamiltonian term
\begin{equation}
    H_{\rm latt} = \sum_{\bf n} \left[ \frac{K}{2} \left( {u_{\bf n}^x}^2 + {u_{\bf n}^y}^2 \right) + \frac{m}{2} \left({v_{\bf n}^x}^2 + {v_{\bf n}^y}^2 \right) \right],
\label{eq:Hla}
\end{equation}
where $K$ is the force constant associated with the $\rm O$ ion displacements, $m$ is the mass of the $\rm O$ ion, and $v^{a}_{\bf n}=d u^{a}_{\bf n}/d t$ $(a=x,y)$ is the velocity.

The total Hamiltonian is the sum of the above terms,
\begin{equation}
    H_{\rm tot} = H_{\rm hop} + H_{\rm el \mbox{-} latt} + H_{\rm latt},
\end{equation}
which results in the electron energy levels,
\begin{equation}\label{eq:level}
    \varepsilon _{l\bf k} = (-1)^{l} \sqrt{4t_{h}^2(\cos k_x + \cos k_y)^2 + \lambda^2 u^2},
\end{equation}
with the band index $l=0, 1$ and ${\bf k}=(k_{x},k_{y})$ in the first Brillouin zone $\Omega_{\rm 1BZ}=\left \{ {\bf k} | \ |k_{x}|+|k_{y}|\leq \pi  \right \}$. The distribution function for the state $|l{\bf k}\rangle$ is represented by $f_{l \bf k}$. A gap $\Delta _{\rm gap}=2\lambda \left | u \right |$ occurs at the boundary of $\Omega_{\rm 1BZ}$. The metallic state with $u=0$ has a Peierls instability with the Fermi surface nesting vector ${\bf Q}=(\pi, \pi)$. Therefore, the CO insulating phase develops, as $|u|$ becomes finite.

The order parameter for the CO state is defined as the ${\bf Q}=(\pi ,\pi )$ component of the charge density modulation at the $M$ ion sites that is,
\begin{equation}
    \delta n = \frac{1}{N^{2}}\sum_{\bf n}e^{i {\bf Q}\cdot {\bf n}}\langle c_{\bf n}^\dag c_{\bf n}\rangle.
\end{equation}
We choose the size of our system $N = 512$. The hopping constant $t_{h} = 0.5$~eV, the electron-lattice coupling constant $\lambda = 0.936$~eV~\AA$^{-1}$, and the force constant $K=0.85$~eV~\AA$^{-2}$ are chosen similar to the values used for perovskite manganites.~\cite{Ahn2000PRB,Ahn2001PRB} The mass of the oxygen ion is $m = 1.66$~meV~ps$^2$~\AA$^{-2}$. While the dynamics of a particular phonon mode directly coupled to the CO is coherent and parameterized by $u$, the rest of phonon system is assumed to be incoherent and play the role of a thermal reservoir to the electron system excited by the optical pump, because the phonon system has a much greater specific heat than the electron system.
To simulate the role as a thermal reservoir, we describe the state of the incoherent phonon system by the Bose-Einstein distribution function $b_{\omega}$ with the temperature fixed at the initial temperature $T_i$ as done in Ref.~\onlinecite{Groeneveld1992PRB,Groeneveld1995PRB}, and consider the scattering between electrons and phonons. The phonon DOS per site $D_{p}( \omega )$ is obtained by modifying the Debye model.
Below the Debye energy $\omega_{D}$, the phonon DOS $D_{p}( \omega )$ is proportional to $\omega^{2}$. Above $\omega_{D}$, a Gaussian function is assumed with the peak at $\omega_{D}$ matched to $D_{p}(\omega )$ of the Debye model,
\begin{equation}
    D_{p}(\omega) = \left\{\begin{matrix}
        \zeta \omega^2 & {\rm for}\ 0\leq\omega\leq\omega_D ,\\
        \zeta \omega_{D}^2 e^{-(\omega - \omega_D)^2 / \eta^2} & {\rm for}\ \omega > \omega_D.
    \end{matrix}\right.
\end{equation}
We choose $\omega _D = 60$~meV and $\eta = 15$~meV, so that the highest phonon energy is similar to the phonon energy for the typical $M$-$\rm O$ bond stretching mode. The total number of phonon modes per site is chosen as 5 to match to the number of longitudinal phonon modes per transition metal ion in perovskite transition metal oxides, which sets $\zeta  = 3.63 \times 10^{-5}$~meV$^{-3}$.


\subsection{Lattice dynamics}

In the model, the coherent lattice distortion parameterized by $u$ is treated classically and follows  Newtonian dynamics. The corresponding potential energy per site is given by
\begin{equation}
    U(u) = \frac{1}{N^{2}} \sum_{l{\bf k}} \varepsilon_{l{\bf k}}(u) f_{l{\bf k}} + K u^2.
\label{eq:potential}
\end{equation}
The Lagrangian per site $\mathcal{L} = m v^{2} - U(u)$ with $v=du/dt$ and the damping lead to the equation for the dynamics of the distortion~$u$,
\begin{equation}
    2m \frac{d ^2u}{d t^2} = -2Ku-\frac{1}{N^{2}}\sum_{l{\bf k}} \frac{\partial \varepsilon _{l{\bf k}}(u)}{\partial u}f_{l{\bf k}} - \gamma \frac{d u}{d t} ,
\end{equation}
where a value of damping constant $\gamma = 9$~meV~ps~{\AA}$^{-2}$ is chosen, so that the decay rate of the oscillation is similar to experiments.~\cite{Beaud2014NATMAT}

\subsection{Electron dynamics}

Dynamics of electrons in the model is governed by the Boltzmann equations that describe electron-electron and electron-phonon scattering. As done in Refs.~\onlinecite{Demsar2003PRL,Rethfeld2002PRB,Groeneveld1992PRB,Groeneveld1995PRB,Ahn2004PRB}, the momentum conservation is integrated out under the approximation of isotropic Debye phonons and electrons with isotropic parabolic dispersion relation. This gives rise to the following equations
\begin{equation}
    \frac{d f_\varepsilon}{d t} = \left( \frac{d f_\varepsilon}{d t} \right )_{\!\!\rm ee}+\left( \frac{d f_\varepsilon}{d t} \right)_{\!\!\rm ep} ,
\end{equation}
where
\begin{eqnarray}
    \left( \frac{d f_\varepsilon}{d t} \right)_{\!\!\rm ee} &=& \frac{K_{\rm ee}}{2} \int \big[ -f_{\varepsilon }f_{{\varepsilon }'}(1-f_{{\varepsilon }''})(1-f_{\varepsilon +{\varepsilon }'-{\varepsilon }''}) \nonumber \\
    && + (1-f_{\varepsilon })(1-f_{{\varepsilon }'})f_{{\varepsilon }''}f_{\varepsilon +{\varepsilon }'-{\varepsilon }''} \big] \nonumber \\
    && \times D_{e}({\varepsilon }')D_{e}({\varepsilon }'')D_{e}(\varepsilon +{\varepsilon }'-{\varepsilon }'') d\varepsilon' d\varepsilon'' 
\end{eqnarray}
represents the electron-electron scattering, and
\begin{eqnarray}
    \left( \frac{d f_\varepsilon}{d t} \right)_{\!\!\rm ep} &=& K_{\rm ep} \int \left\{\big[ f_{\varepsilon+\omega}(1-f_{\varepsilon })(b_{\omega }+1) \right. \nonumber \\
    &&- f_{\varepsilon }(1-f_{\varepsilon +\omega })b_{\omega }\big] D_{p}(\omega)D_{e}(\varepsilon+\omega) \nonumber \\
    && + \big[f_{\varepsilon -\omega }(1-f_{\varepsilon })b_{\omega }-f_{\varepsilon }(1-f_{\varepsilon -\omega })(b_{\omega }+1) \big] \nonumber \\
    && \left.\times D_{p}(\omega)D_{e}(\varepsilon-\omega) \right\}d\omega 
\end{eqnarray}
represents the electron-phonon scattering, in terms of electron and phonon distribution functions, $f_{\varepsilon }$ and $b_{\omega }$, and corresponding DOS, $D_{e}(\varepsilon )$ and $D_{p}(\omega )$. 
The number of energy bins is chosen as $N_e = 2400$, which results in an energy bin size of about 1.7~meV. The constants for the electron-electron and the electron-phonon scattering are $K_{\rm ee} = 1953$~eV~ps$^{-1}$ and $K_{\rm ep} = 0.2325$~eV~ps$^{-1}$, chosen with the same order of magnitude as the values used in Refs.~\onlinecite{Demsar2003PRL,Ahn2004PRB}.
\subsection{Approximations used}
We list some of the approximations chosen for the model and discuss why they are reasonable. In the simulations, the electron DOS plays a dominant role in dynamics. Electron hopping amplitudes beyond the nearest neighbors are not only small, but also have a negligible effect on the electron DOS, which justifies the approximation of including only the nearest neighbor hopping. An approximation has been also made for the effect of the optical pump. The main focus of the simulations is the dynamics \textit{after} the optical pump, not \textit{during} the optical pump. Further, the typical width of the optical pulse, $\sim$10~fs, is much shorter than the period of coherent oscillation, $\sim$500~fs. Therefore, the dynamics during the optical pump is irrelevant for the simulation and we approximate the effect of the optical pump as an instantaneous electronic excitation,~\cite{Ahn2004PRB,Demsar2003PRL} as described in Sec.~\ref{sec:results}~A.

Finally, all phonon modes, except one primary coherent distortion mode parameterized by $u$, have no memory of the phonons emitted or absorbed by electrons, and are treated as a thermal reservoir at a fixed temperature. The effect of dynamic incoherent phonon distribution is expected to be small, because the phonons have a much greater specific heat than the electrons.
Excitations of other coherent phonon modes coupled to the primary coherent phonon mode could be incorporated in the model by including anharmonic coupling between various coherent phonon modes, as postulated for perovskite manganites.~\cite{Beaud2014NATMAT} 

With these reasonable approximations, we capture essential features of CO and its photoinduced dynamics in a simple model, and obtain results which could spur future experiments. The model also provides a computational framework, upon which more realistic models could be built.
\section{Results}\label{sec:results}

\subsection{Equilibrium states and excitations by optical pump}

Before presenting the results for the nonequilibrium dynamics, we discuss the equilibrium properties of the system and the effects of the optical pump. To ensure consistency, the dynamics simulation itself is used to obtain the equilibrium states $f_{\varepsilon }^{\rm eq}$ and $u_{\rm eq}$, which show a second order phase transition with a critical temperature of $T_{c}\approx 217$~K and $u_{\rm eq}(T\approx 0)=0.035$~{\AA}.

As mentioned in Sec.~\ref{sec:model}~D, the effect of the optical pump is considered as an instantaneous electron excitation. Therefore, the distribution function for the upper and lower bands at $t=0$, right after the optical pump, is altered from the equilibrium distribution $f_{\varepsilon }^{\rm eq}$ by a Gaussian function,
\begin{equation}\label{eq:pump}
     f_{\varepsilon }(t=0) = f_{\varepsilon }^{\rm eq} \pm \mathit{\delta f}\exp\left [  -\frac{(2\varepsilon \mp E_{\rm photon})^{2}}{8W^{2}}\right ] ,
\end{equation}
where $E_{\rm photon}$ is the median photon energy in the optical pump, and $\mathit{\delta f}$ is the maximum change in the distribution function. The fluence per site $F$ of the optical pump is calculated as the change in electronic energy at $t=0$. For most results in this paper, we take an initial temperature of $T_{i}=135$~K, for which the equilibrium distortion, order parameter, and CO gap are $u_{\rm eq}= 0.031$~\AA, $\delta n_{\rm eq}=0.056$, and $\Delta_{\rm gap}=58.9$~meV, respectively. The width of the pump beam is fixed as $W=0.02$~eV for most simulations.

\subsection{Nonequilibrium dynamics induced by photons with $E_{\rm photon}\gg \Delta _{\rm gap}$}\label{ssec:induced}

\begin{figure}
    \centering
    \includegraphics[width=1.0\hsize,clip]{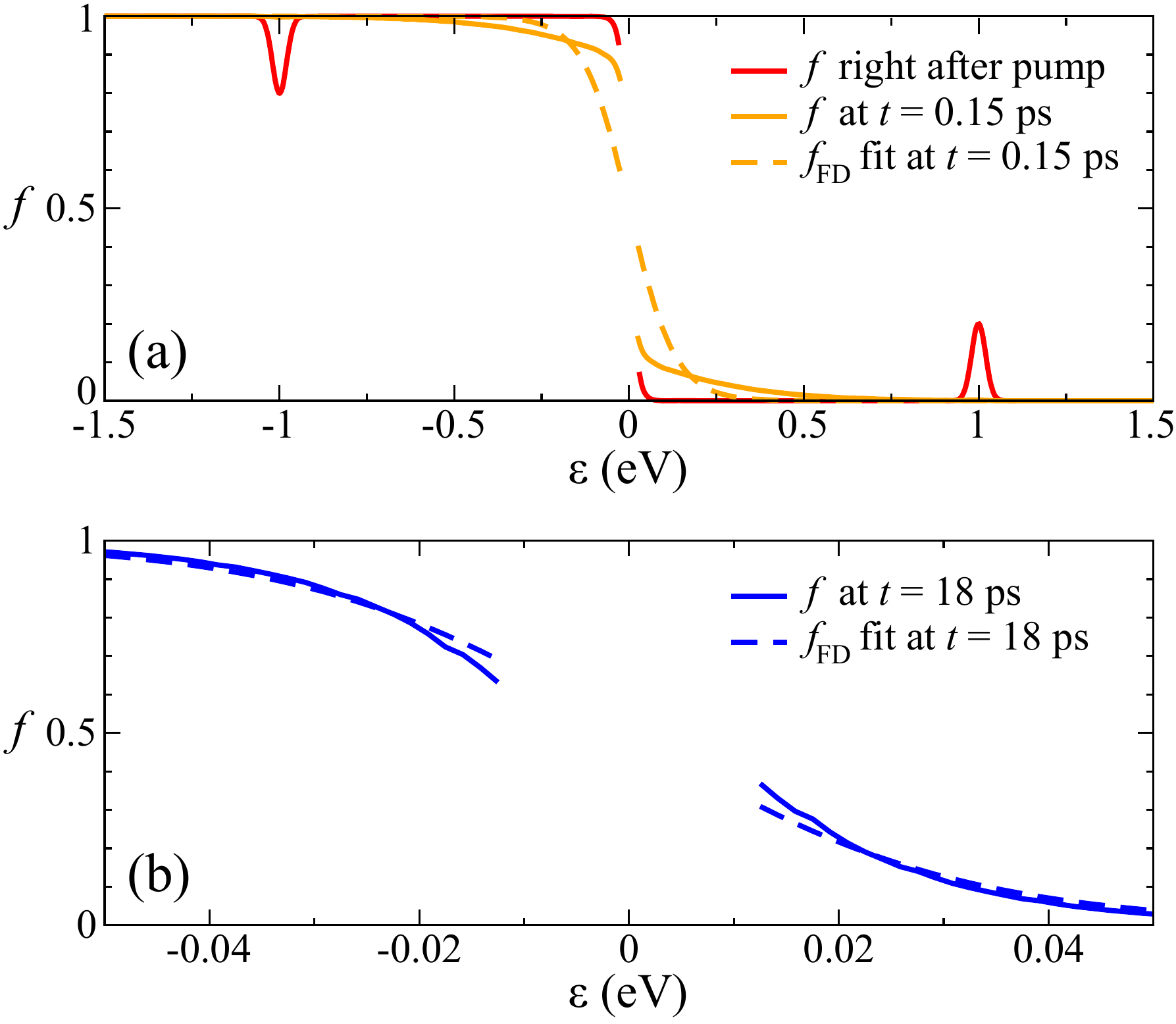}
    \caption{
    Examples for the evolution of the distribution function $f$ versus energy $\varepsilon$ for $E_{\rm photon} \gg \Delta_{\rm gap}$. (a) The red line shows the  distribution function right after the optical pump, $t=0$; the solid orange line represents the distribution function at $t=0.15$~ps. The Fermi-Dirac distribution at the corresponding effective temperature $T_{\rm eff}=904$~K, defined by Eq.~(\ref{eq:intmatch}), at $t=0.15$~ps is shown in a dashed orange line. (b) The solid blue line represents the electron distribution at $t=18$~ps during the gap reopening. The Fermi-Dirac distribution with the corresponding effective temperature $T_{\rm eff}=177$~K at $t=18$~ps is shown in a dashed blue line.
    }\label{fig:distrib}
\end{figure}

\begin{figure}
    \centering
    \includegraphics[width=1.0\hsize,clip]{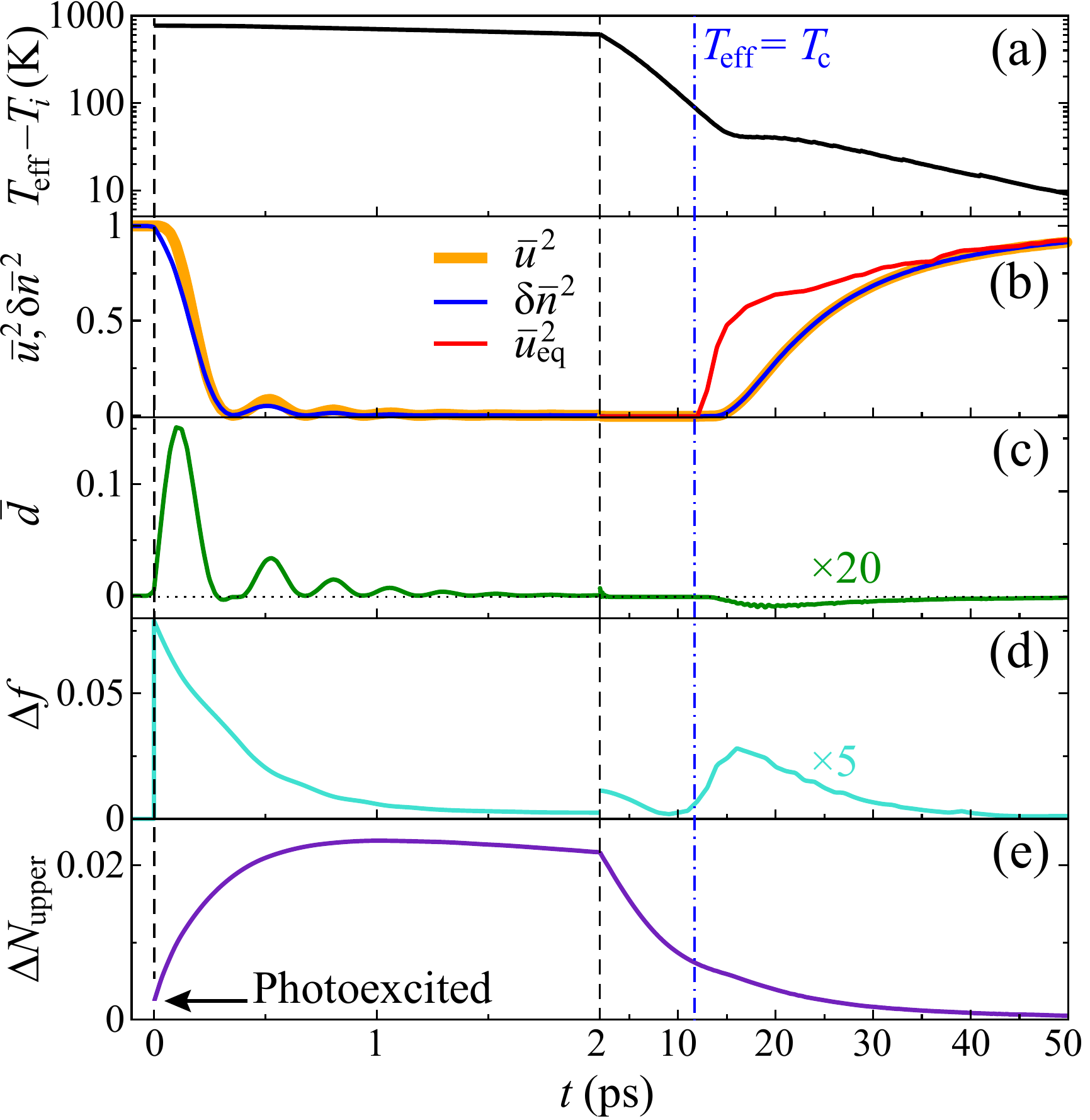}
    \caption{
    Nonequilibrium dynamics for $E_{\rm photon} \gg \Delta_{\rm gap}$ and $F=4.38~{\rm meV/site} > F_{c}$, the critical fluence for photoinduced insulator-metal transition. The time scales up to 2~ps and between 2~ps and 50~ps are chosen differently to reveal features more clearly. (a) $T_{\rm eff}-T_{i}$, the difference between the effective temperature of the electron system defined by Eq.~(\ref{eq:intmatch}) and the initial temperature before the pump, (b) the square of periodic lattice distortion $\bar u^{2}$, the square of electronic order parameter $\delta \bar n^{2}$, and the square of equilibrium distortion $\bar u^{2}_{\rm eq}$ at $T_{\rm eff}(t)$, normalized to their values before the optical pump, (c) $\bar d=\bar u^{2}-\delta \bar n^{2}$, which parameterizes the decoupling between the CO and the periodic lattice distortion, (d) $\Delta f$, the average deviation of the electron distribution function $f(\varepsilon)$ from the Fermi-Dirac distribution function $f_{\rm FD}(\varepsilon ,T_{\rm eff})$, defined by Eq.~(\ref{eq:deltaf}), (e) $\Delta N_{\rm upper}$ defined by Eqs.~(\ref{eq:Nupper}) and (\ref{eq:deltaNupper}), that is, the number of excess electrons per site in the upper band with respect to the equilibrium state before the optical pump. The horizontal arrow indicates the number of electrons excited by the optical pump. For clarity, $\bar{d}$ and $\Delta f$ between 2~ps and 50~ps are multiplied by constant factors indicated in the figure.
    }\label{fig:set}
\end{figure}

Since the early-time dynamics and the energy efficiency of melting the CO depend sensitively on the photon energy, the results for $E_{\rm photon}\gg \Delta _{\rm gap}$ and $E_{\rm photon} = \Delta _{\rm gap}$ are presented separately in this and the next subsections. The results for $E_{\rm photon}=2$~eV, much greater than $\Delta_{\rm gap}=58.9$~meV, and fluences large enough to melt the CO are presented in Figs.~\ref{fig:distrib} and~\ref{fig:set}. In Fig.~\ref{fig:distrib}, the electron distribution functions for selected times are shown to demonstrate the evolution of $f(\varepsilon )$. Video simulations of $f(\varepsilon ,t)$ for the $E_{\rm photon} \gg \Delta_{\rm gap}$ and $E_{\rm photon} = \Delta_{\rm gap}$ cases are provided in the supplementary material. In Fig.~\ref{fig:set}, the evolution of various quantities are shown. To reveal the fast early dynamics and slow late dynamics in the same figure, the dynamics during $-0.1$-$2$~ps and $2$-$50$~ps are displayed in different time scales. To parameterize the energy of the electron system at time $t$, the effective temperature $T_{\rm eff}(t)$ in the nonequilibrium state is defined by matching the total energy between the actual and the Fermi-Dirac distributions, that is,
\begin{equation}\label{eq:intmatch}
    \int_{-\infty}^\infty \varepsilon f(\varepsilon,t) D_{e}(\varepsilon,t) d\varepsilon = \int_{-\infty}^\infty \varepsilon f_{\rm FD}(\varepsilon,T_{\rm eff}(t)) D_e(\varepsilon,t) d\varepsilon ,
\end{equation}
where $f_{\rm FD}(\varepsilon ,T_{\rm eff}(t))$ is the Fermi-Dirac distribution function with the temperature $T_{\rm eff}(t)$ and the chemical potential zero. Figure~\ref{fig:set}(a) shows the difference between $T_{\rm eff}$ and the initial temperature $T_{i}$ before the pump, which is a measure of the excess energy in the electron system. The effective electron temperature $T_{\rm eff}$ increases to 907~K right after the optical pump. The semilogarithmic plot of $T_{\rm eff}-T_{i}$ versus time $t$ reveals three distinct exponential decay rates, $r = 0.202$~$\rm ps^{-1}$ up to around 16~ps, $r = 0.013$~$\rm ps^{-1}$ between 16~ps and 21~ps, and $r = 0.053$~$\rm ps^{-1}$ after around 21~ps, which correspond to three stages of the relaxation process, that is, stages of CO melting, CO gap reopening, and thermal relaxation. Such multistage relaxation has been observed in CO or CDW materials.~\cite{Demsar1999PRL} We now discuss these different stages in more detail.


As shown in Fig.~\ref{fig:distrib}(a), initial electron-hole excitations for $E_{\rm photon}\gg \Delta_{\rm gap}$ occur far away from the CO gap, but fast electron-electron scattering removes the Gaussian peak features at $\varepsilon=\pm E_{\rm photon}/2$ within 0.15~ps, initiating the stage of CO melting. As mentioned in Sec.~\ref{sec:model}, the CO accompanies periodic lattice distortions. Such electronic and lattice modulations would produce superlattice peaks in x-ray and neutron scattering. Their normalized intensities are approximately squares of the displacement $u$ or the CO density $\delta n$ normalized to the equilibrium values at temperature $T_{i}$ before the optical pump,
\begin{eqnarray}
    \bar{u}^{2}(t)&=&\left [ u(t)/u_{\rm eq}(T_{i}) \right ]^{2},  \nonumber \\
    \delta \bar{n}^{2}(t)&=&\left [\delta n(t)/\delta n_{\rm eq}(T_{i}) \right ]^{2}.
\label{eq:normalsq}
\end{eqnarray}
In equilibrium, $u$ and $\delta n$ are directly related to each other via
\begin{equation}
    u_{\rm eq} = \frac{\lambda}{2K} \delta n_{\rm eq}.
\end{equation}
Therefore, we define
\begin{equation}
    \bar{d}(t) = \bar{u}^{2}(t)-\delta \bar{n}^{2}(t)
\end{equation}
to characterize the decoupling between the CO and periodic lattice distortion in nonequilibrium. Figure~\ref{fig:set}(b) shows that substantial electron-hole excitations near the gap created by the electron-electron scattering reduce the order parameter $\delta n$ and initiate the coherent oscillation in $u$, which damps out by around 1~ps. The result further reveals a difference between $\bar{u}^{2}$ and $\delta \bar{n}^{2}$, up to approximately 15\% at $t \approx 0.13$~ps, as shown more clearly for $\bar{d} = \bar{u}^{2} - \delta \bar{n}^{2}$ in Fig.~\ref{fig:set}(c), which indicates a partial decoupling of the electrons and lattice distortions. The $\rm O$ ion has about thirty thousand times greater mass than an electron, which results in lattice dynamics lagging behind the electron dynamics and $\bar d > 0$. The oscillation amplitude of the normalized lattice distortion is larger than that of the normalized electronic order parameter for the same reason. The average difference $\Delta f(t)$ between $f(\varepsilon,t)$ and $f_{\rm FD}(\varepsilon, T_{\rm eff}(t))$, calculated according to
\begin{equation}\label{eq:deltaf}
    \Delta f(t) = \sqrt{\int_{-\infty }^\infty  \left [  f(\varepsilon,t)-f_{\rm FD}(\varepsilon,T_{\rm eff}(t)) \right ]^2 D_{e}(\varepsilon,t) d\varepsilon },
\end{equation}
is shown in Fig.~\ref{fig:set}(d), which indicates that the electronic state deviates substantially from the Fermi-Dirac distribution during the CO melting. To track the transfer of electrons between the upper and the lower bands, we calculate the number of electrons per site in the upper band at time~$t$,
\begin{equation}\label{eq:Nupper}
    N_{\rm upper}(t) = \int_0^\infty f(\varepsilon,t) D_{e}(\varepsilon,t) d\varepsilon,
\end{equation}
and find the change from the number before the optical pump,
\begin{equation}\label{eq:deltaNupper}
    \Delta N_{\rm upper}(t) = N_{\rm upper}(t)-N_{\rm upper}(t<0),
\end{equation}
shown in Fig.~\ref{fig:set}(e). The number of photoexcited electrons in the upper band is 0.002 per site, as indicated by a horizontal arrow in Fig.~\ref{fig:set}(e), while the number of electrons excited through the subsequent thermalization up to $\sim 1$~ps is 0.021 per site, an order of magnitude greater, because many \textit{low} energy electrons are excited near the gap as photoexcited \textit{high} energy electrons decay through the energy-conserving electron-electron scattering.

As the effective electron temperature $T_{\rm eff}$ drops below $T_{c}$ around $t=12$~ps, indicated by the vertical dot-dashed blue line in Fig.~\ref{fig:set}, the electron system enters the stage of CO gap reopening, and loses the $internal$ equilibrium up to approximately $t=30$~ps. Figure~\ref{fig:set}(b) shows that the squares of electronic order parameter and periodic lattice distortion, $\delta \bar{n}^{2}$ and $\bar{u}^{2}$, increase from zero. The square of the normalized lattice distortion that the system would have, if the system is in the equilibrium state at $T_{\rm eff}$,
\begin{equation}
    \bar {u}_{\rm eq}^{2}(t)=\left [ \frac{u_{\rm eq}(T_{\rm eff}(t))}{u_{\rm eq}(T_{i})} \right]^{2} ,
\end{equation}
is also shown in Fig.~\ref{fig:set}(b) for $t>2$~ps. The strong reduction of the normalized actual distortion $\bar u(t)$ compared to the normalized equilibrium distortion $\bar u_{\rm eq}(t)$ clearly shows the effect of nonequilibrium dynamics. The electronic ordering precedes the lattice ordering again and therefore $\bar{d} =  \bar{u}^{2} - \delta \bar{n}^{2}<0$ [Fig.~\ref{fig:set}(c)]. Furthermore, rapid opening of the gap pushes electron and hole energies up, which causes a very slow decay of $T_{\rm eff}$ [Fig.~\ref{fig:set}(a)], a substantial deviation of $f(\varepsilon)$ from $f_{\rm FD}(\varepsilon)$ near the gap [Fig.~\ref{fig:distrib}(b)], and enhanced $\Delta f$ [Fig.~\ref{fig:set}(d)]. We discuss this in more detail in Sec.~\ref{ssec:nonthermal}.

Finally, the stage after around 30~ps is characterized as the thermal relaxation stage, because the electron system and the periodic lattice distortion gradually approach the initial state before the optical pump, while maintaining $internal$ equilibrium between them.

\begin{figure}
    \centering
    \includegraphics[width=1.0\hsize,clip]{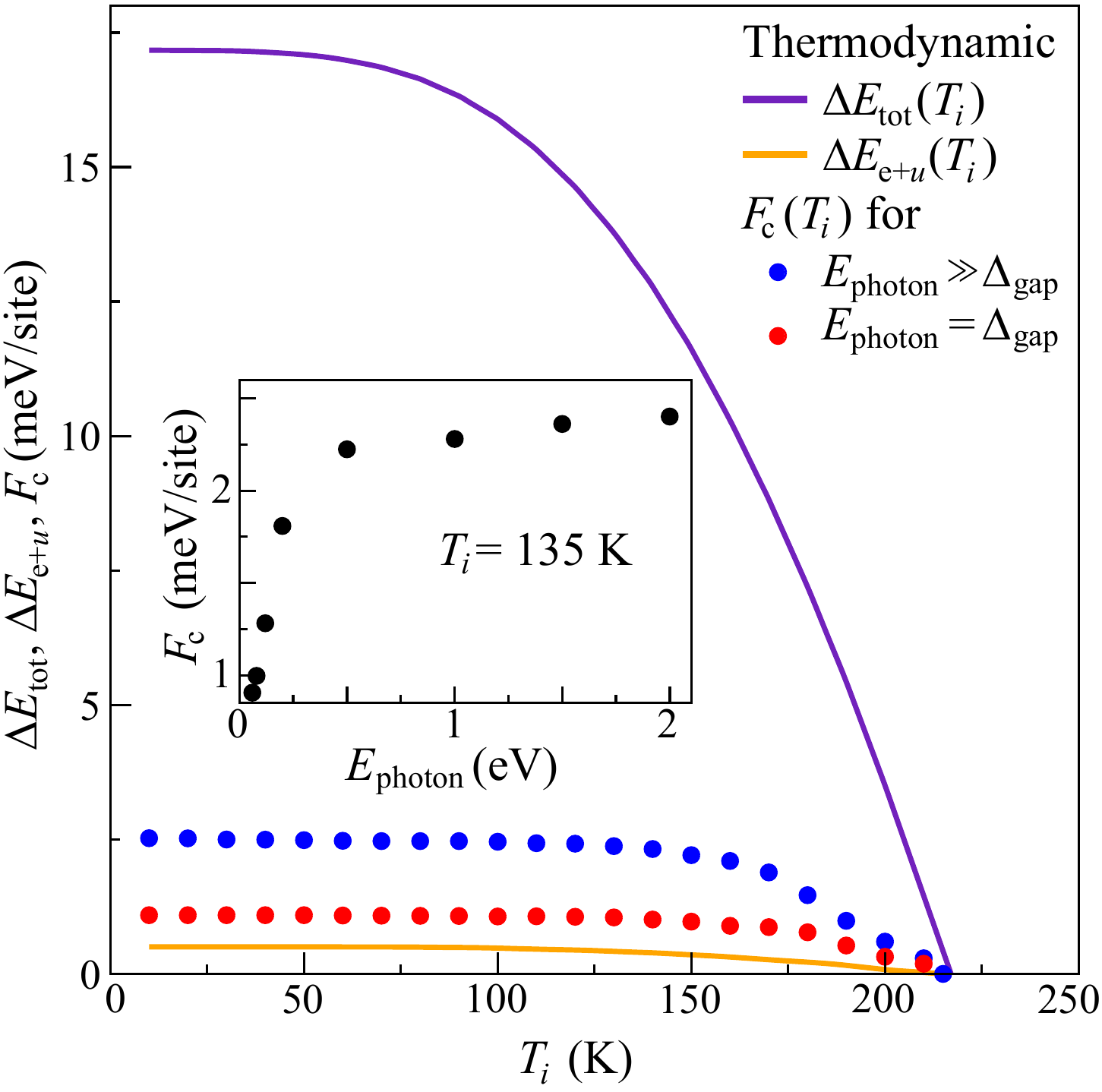}
    \caption{
    Comparison between the energies required for the thermodynamic and photoinduced insulator-metal transitions. The purple [orange] line represents the thermodynamic melting energy $\Delta E_{\rm tot}(T_{i})$ [$\Delta E_{{\rm e}+u}(T_{i})$] with [without] incoherent phonons, that is, the energy needed to thermodynamically heat the system including [excluding] incoherent phonons from $T_{i}$ to $T_{c}$. The blue and red dots in the main panel represent the critical fluence $F_{c}(T_{i})$ for the photoinduced insulator-metal transition by  the pump beams with $E_{\rm photon} \gg \Delta_{\rm gap}$ and $E_{\rm photon} = \Delta_{\rm gap}$, respectively. The inset shows $F_{c}$ versus $E_{\rm photon}$ at a fixed initial temperature $T_{i}=135$~K.
    }\label{fig:thermo}
\end{figure}

The critical fluence $F_{c}$ for the insulator-metal transition versus the initial temperature $T_{i}$ before the pump for $E_{\rm photon}\gg \Delta_{\rm gap}$ is shown in blue dots in Fig.~\ref{fig:thermo}. To compare the photoinduced and thermodynamic insulator-metal transitions, we calculate the thermodynamic CO melting energy $\Delta E_{\rm tot}(T_{i})$ at temperature $T_{i}<T_{c}$ for the whole system including the phonon thermal reservoir according to
\begin{equation}\label{eq:Etot}
    \Delta E_{\rm tot}(T_{i})=E_{\rm tot}(T_{c})-E_{\rm tot}(T_{i}),
\end{equation}
where
\begin{eqnarray}
   E_{\rm tot}(T)&=&\int_{-\infty }^{\infty }\varepsilon f_{\rm FD}(\varepsilon ,T)D_{e}(\varepsilon ,u_{\rm eq}(T))d\varepsilon +Ku_{\rm eq}^{2}(T)  \nonumber \\
   &+&\int_{0}^{\infty }\omega b_{\omega}(T)D_{p}(\omega)d\omega.
\label{eq:Etot2}
\end{eqnarray}
The result shown in purple line in Fig.~\ref{fig:thermo} indicates that the energy required for the photoinduced phase transition $F_{c}(T_{i})$ is substantially lower than the energy required for the thermodynamic phase transition $\Delta E_{\rm tot}(T_{i})$ for the model system, because for the photoinduced transitions there is insufficient time to heat the incoherent phonons. We also find the thermodynamic melting energy \textit{without} the incoherent phonons $\Delta E_{{\rm e}+u}(T_{i})$ shown in orange line in Fig.~\ref{fig:thermo}, by excluding the last term in Eq.~(\ref{eq:Etot2}) and calculating the difference between $T_{c}$ and $T_{i}$. The critical fluence $F_{c}(T_{i})$ is greater than $\Delta E_{{\rm e}+u}(T_{i})$, indicating that a part of the energy initially deposited to the electron system leaks to the phonon thermal reservoir before the high energy electron and hole pairs cascade down to the states near the gap and initiate the insulator-metal transition.

\subsection{Nonequilibrium dynamics induced by photons with $E_{\rm photon} = \Delta_{\rm gap}$}

\begin{figure}
    \centering
    \includegraphics[width=1.0\hsize,clip]{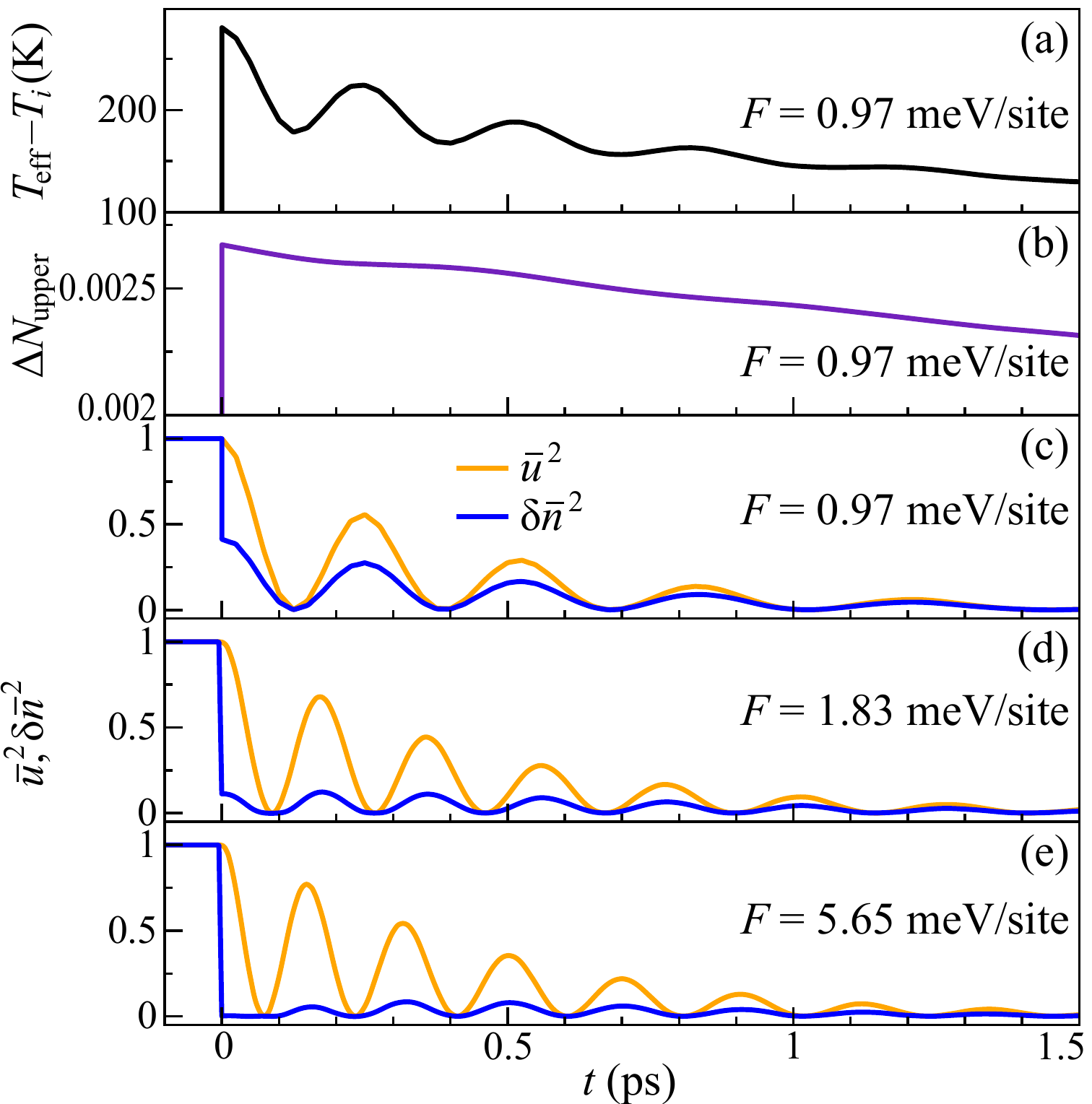}
    \caption{
    Nonequilibrium dynamics for $E_{\rm photon} = \Delta_{\rm gap}$. (a) The difference between the effective electron temperature $T_{\rm eff}$ and the initial temperature $T_{i}$ before the pump. (b) $\Delta N_{\rm upper}$, the number of excess electrons in the upper band with respect to the equilibrium state before the pump. (c), (d), and (e): Normalized squares of the periodic lattice distortion and electronic order parameter, $\bar{u}^{2}$ and $\delta \bar{n}^{2}$, defined in Eq.~(\ref{eq:normalsq}) versus time $t$. The fluence of the pump beams are (c) $F=0.97$~meV/site, (d) 1.83~meV/site, and (e) 5.65~meV/site, all above the critical fluence $F_{c}=0.91$~meV/site. For (e), we use $W=0.06$~eV as the width of the Gaussian peak in Eq.~(\ref{eq:pump}).
    }\label{fig:fluence}
\end{figure}

In this subsection, the results of the simulations with $E_{\rm photon} = \Delta_{\rm gap}$ are presented, particularly before 1.5~ps when the dynamics shows a behavior different from the case of high photon energy $E_{\rm photon}\gg \Delta_{\rm gap}$. The dynamics of the square of the normalized distortion $\bar{u}^{2}$ and the square of the normalized order parameter $\delta \bar{n}^{2}$ are shown in Figs.~\ref{fig:fluence}(c)-\ref{fig:fluence}(e) for three values of the fluence $F=0.97$, 1.83, and 5.65~meV/site, all above the critical fluence $F_{\rm c}=0.91$~meV/site. At $t = 0$, while $\bar{u}^{2}$ still decreases continuously, the electronic parameter $\delta \bar{n}^{2}$ jumps abruptly by the amount that increases with the fluence $F$. This jump in $\delta \bar{n}^{2}$ occurs because the electrons with energies right at the gap, which are relevant to the CO, are directly excited by the optical pump. Figure~\ref{fig:fluence}(e) shows that the electronic order virtually vanishes and remains close to zero for a high enough fluence with $E_{\rm photon}$ close to $\Delta_{\rm gap}$. The energy of the electrons and holes excited by the optical pump near the gap is strongly coupled to $\bar{u}$ and gives rise to oscillating effective electron temperature $T_{\rm eff}$, as shown in Fig.~\ref{fig:fluence}(a) for $F=0.97$~meV/site. With the low energy of the photoexcited electrons, the initial electron thermalization reduces $\Delta N_{\rm upper}$, as indicated in Fig.~\ref{fig:fluence}(b), very different from $E_{\rm photon} \gg \Delta_{\rm gap}$ case shown in Fig.~\ref{fig:set}(e). Figures~\ref{fig:fluence}(c)-\ref{fig:fluence}(e) also show that the period of oscillation depends sensitively on time and the fluence, which will be analyzed in more detail in the next subsection.

Red dots in Fig.~\ref{fig:thermo} show the critical fluence $F_{c}$ versus the initial temperature $T_{i}$ for $E_{\rm photon} = \Delta_{\rm gap}$ and the inset in Fig.~\ref{fig:thermo} displays $F_{c}$ versus $E_{\rm photon}$ at $T_{i}=135$~K. As the photon energy decreases from $E_{\rm photon} \gg \Delta_{\rm gap}$ to $E_{\rm photon} = \Delta_{\rm gap}$, the critical fluence $F_{c}(T_{i})$ reduces by about $60$\%, toward the thermodynamic melting energy $\Delta E_{{\rm e}+u}(T_{i})$ without incoherent phonons. The melting of the CO is greatly facilitated by exciting the electrons close to the gap, because photons in the optical pump directly alter the CO and more energy is used for the CO melting.

\subsection{Dynamics of energy landscape and coherent oscillation frequency}\label{ssec:landscape}

The energy landscape plays an important role in both thermodynamic and photoinduced phase transitions. We therefore calculate the dynamic energy landscape $U(u,t)$ according to
\begin{equation}\label{eq:landscape}
    U(u,t)=\frac{1}{N^{2}}\sum_{l{\bf k}}\varepsilon _{l {\bf k}}(u)f_{l{\bf k}}(t)+Ku^{2},
\end{equation}
where the first term, the electron energy summed over the occupation, represents the electron-lattice coupling, and the second term represents the vibrational potential energy from ion-ion interactions.
In the first term, the electron distribution $f_{l{\bf k}}$ in the band and momentum indices is independent of the distortion $u$, because $u$ is varied adiabatically. Strikingly different early-time energy landscape dynamics are found for different photon energies, as shown in Fig.~\ref{fig:landsc}. Video simulations of $U(u,t)$ are provided in the supplementary material. Figures~\ref{fig:landsc}(a) and~\ref{fig:landsc}(c) display the results for $E_{\rm photon} \gg \Delta_{\rm gap}$. The energy landscape right after the optical pump at $t=0$ [red line in Fig.~\ref{fig:landsc}(a)] is close to a vertical shift of the energy landscape before the pump at $t<0$ (black line), because the electronic excitations far away from the gap do not couple strongly to the periodic distortion $u$. Subsequently, the energy landscape changes from a double-well (red and orange lines) to a single-well potential (green and cyan lines), and as the effective temperature $T_{\rm eff}$ drops below $T_{\rm c}$ the energy landscape becomes double-well again (dark blue and purple lines). Figure~\ref{fig:landsc}(c) shows the full energy landscape dynamics in $U$-$u$-$t$ space, along with the dynamics of the distortion $u$. Energy landscape changes from a double-well to a single-well during the first oscillation in $u$, resulting in a slow first oscillation, as discussed in more detail later in this subsection.
\begin{figure}
    \centering
    \includegraphics[width=1.0\hsize,clip]{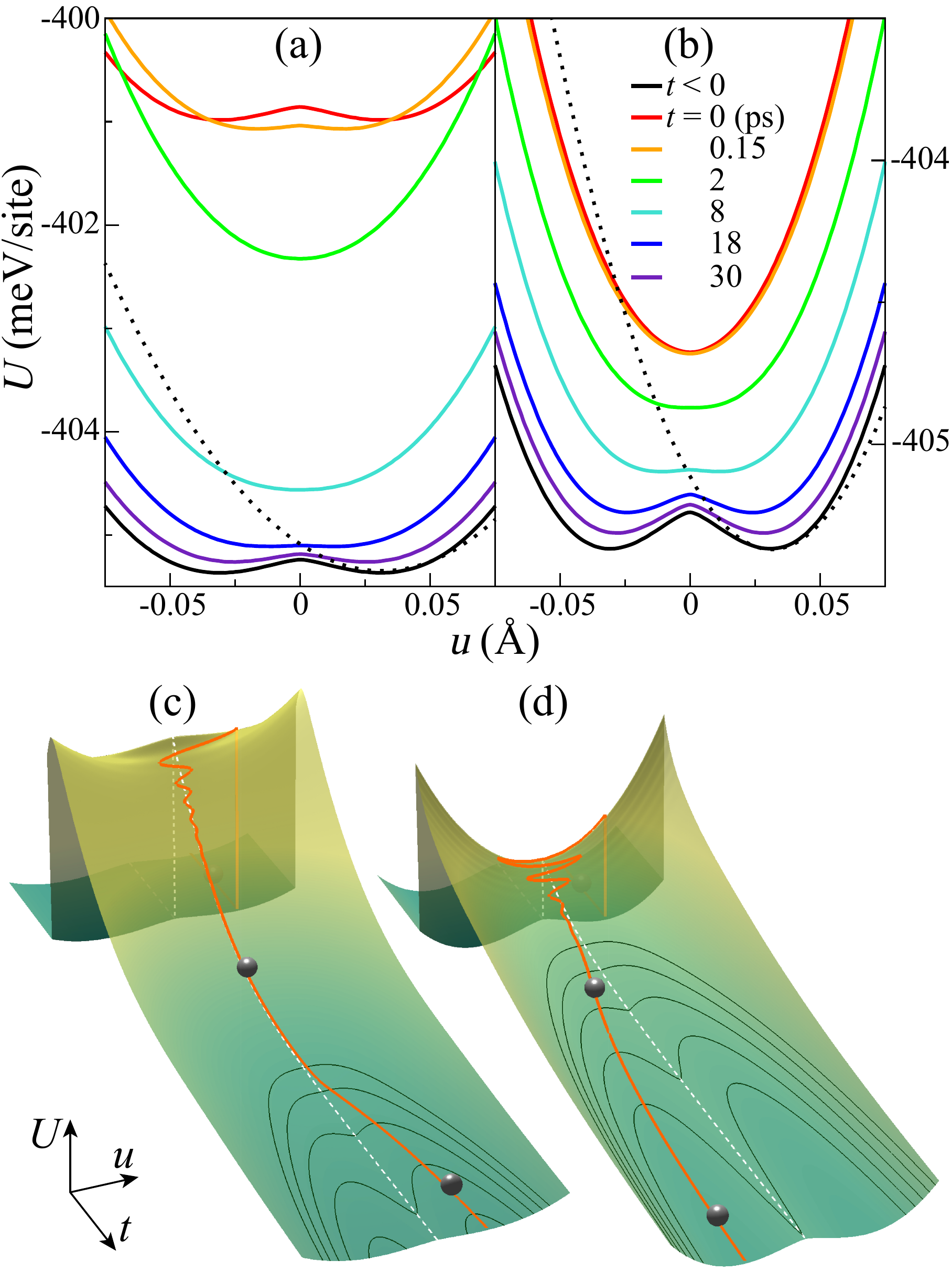}
    \caption{
    Energy landscape dynamics initiated by the optical pump with $E_{\rm photon}=2~\rm eV \gg \Delta_{\rm gap}$ and $F=4.38~\rm meV/site$ for (a) and (c), and with $E_{\rm photon} = \Delta _{\rm gap}$ and $F=0.97~\rm meV/site$ for (b) and (d). In (a) and (b), the dotted black lines represent the harmonic energy near the equilibrium distortion before the optical pump. In (c) and (d), the orange and green lines represent $u(t)$ and equal-energy lines respectively, with the axis ranges of $-5$~ps~$<t<$~25~ps and $-0.06$~{\AA}~$<u<$~$0.06$~{\AA}.
    }\label{fig:landsc}
\end{figure}

The energy landscape dynamics for the $E_{\rm photon}=\Delta_{\rm gap}$ case in Figs.~\ref{fig:landsc}(b) and~\ref{fig:landsc}(d) show a behavior very different from the $E_{\rm photon}\gg\Delta_{\rm gap}$ case, particularly during $t<3$~ps. With the CO state directly destroyed by the optical pump, the energy landscape right after the optical pump [red line in Fig.~\ref{fig:landsc}(b)] already has a metallic single-well potential.  Comparison between $E_{\rm photon}\gg \Delta_{\rm gap}$ case and $E_{\rm photon}= \Delta_{\rm gap}$ case in Fig.~\ref{fig:landsc} reveals that, when the pump energy is tuned at the gap, the change in the shape of the energy landscape occurs in the time scale of the pump pulse width, resulting in much faster and more energy efficient melting of the CO phase, which could be important in using such phenomena for ultrafast switching devices.

\begin{figure}
    \centering
    \includegraphics[width=1.0\hsize,clip]{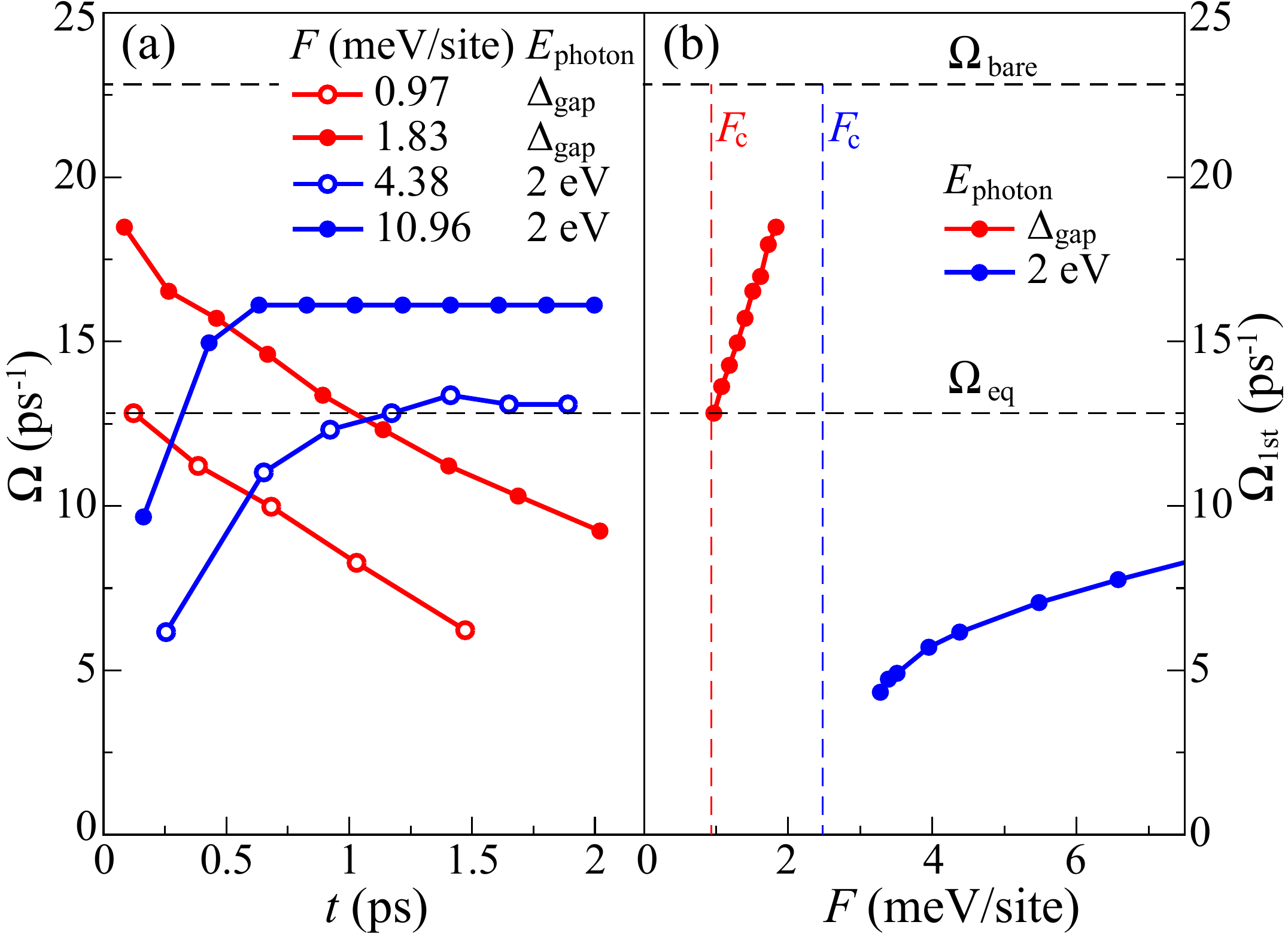}
    \caption{
    (a) Angular frequency of coherent oscillation $\Omega $ versus time $t$, (b) the first angular frequency $\Omega_{\rm 1st}$ from the first half-oscillation versus the fluence of the optical pump $F$ for $E_{\rm photon} = \Delta_{\rm gap}$ and $E_{\rm photon} = 2~{\rm eV} \gg \Delta_{\rm gap}$. Lines connecting symbols are guides for eyes. The horizontal dashed black lines represents the bare angular frequency $\Omega_{\rm bare} = \sqrt{K/m}$ without electron-lattice coupling and the equilibrium angular frequency $\Omega_{\rm eq}$ before the optical pump. The vertical dashed red and blue lines in (b) indicate the critical fluence $F_{c}$ for $E_{\rm photon}=\Delta_{\rm gap}$ and 2~eV, respectively.
    }\label{fig:freq}
\end{figure}

Energy landscape dynamics can be experimentally observed through the time-dependent frequency of coherent oscillations. To analyze the correlation between the energy landscape and the frequency of the oscillation for the model, we first find the angular frequency $\Omega =\pi /(t_{n+1}-t_{n})$ versus time $t=(t_{n+1}+t_{n})/2$, where $t_{n}$ is the time for the $n$-th local maximum of $\bar{u}^{2}(t)$. Figure~\ref{fig:freq}(a) displays the results for two cases of $E_{\rm photon}=\Delta_{\rm gap}$ [cases of Figs.~\ref{fig:fluence}(c) and \ref{fig:fluence}(d)] and two cases of $E_{\rm photon} \gg \Delta_{\rm gap}$ (including the case in Fig.~\ref{fig:set}).
For comparison, Fig.~\ref{fig:freq} also shows the bare angular frequency $\Omega_{\rm bare}$ without electron-lattice coupling and the equilibrium angular frequency $\Omega_{\rm eq}$ for the equilibrium double-well potential before the pump [see dotted black lines in Figs.~\ref{fig:landsc}(a) and \ref{fig:landsc}(b)]. The results reflect the rapidly changing energy landscape, as the excited electrons and holes redistribute in ways that depend on the photon energy and the fluence. For $E_{\rm photon}=\Delta_{\rm gap}$ shown in red symbols in Fig.~\ref{fig:freq}(a), instantaneous melting of CO leads to $\Omega$ either close to or higher than $\Omega_{\rm eq}$ right after the optical pump, depending on whether $F\approx F_{c}$ or $F\gg F_{c}$.
When the energy landscape is about to change from a single-well to a double-well around 2~ps, the energy landscape becomes highly anharmonic with the flat bottom of potential well [see the curve for $t=2$~ps in Fig.~\ref{fig:landsc}(b)], reflected in a small $\Omega < \Omega_{\rm eq}$ around 2~ps in Fig.~\ref{fig:freq}(a). In contrast, for $E_{\rm photon} \gg \Delta_{\rm gap}$ shown in blue symbols in Fig.~\ref{fig:freq}(a), the situation is reversed. The energy landscape starts out with a double-well that turns into a single-well, reflected in a small $\Omega < \Omega_{\rm eq}$ right after the pump. The angular frequency $\Omega$ increases after melting of CO and remains almost constant for $t=1\sim2$~ps because the energy landscape remains single-well till much later $t\approx 12$~ps.

From the first half-oscillation we take the initial angular frequency $\Omega_{\rm 1st}$ and plot with respect to the fluence $F$ in Fig.~\ref{fig:freq}(b) for $E_{\rm photon}=\Delta _{\rm gap}$ and $E_{\rm photon}=2$~eV $\gg \Delta _{\rm gap}$. The corresponding critical fluences $F_{c}$ are also shown in vertical dashed lines. Much faster increase in $\Omega_{\rm 1st}$ with the fluence for $E_{\rm photon}=\Delta _{\rm gap}$ reflects the dominant effect of the states near the gap on the energy landscape. Transient stiffening of the phonon mode by the optical pump has been identified in a CDW phase of ${\rm CeTe_{3}}$ (Ref.~\onlinecite{Han2012PRB}), and is consistent with the Kohn anomaly,~\cite{GrunerBook,Woll1962PR} that is a softening of the phonon mode responsible for the CO or CDW when approaching $T_{c}$ from above.

\subsection{Nonthermal electron distribution and electron energy relaxation}\label{ssec:nonthermal}
Nonthermal electron distribution could give rise to dynamic behaviors significantly different from the predictions of models based on thermal electron distribution, such as the absence of divergent electron relaxation time at low temperatures in metals that contradicts the prediction from the two-temperature model.~\cite{Groeneveld1992PRB,*Groeneveld1995PRB,Ahn2004PRB,Allen1987PRL,Kabanov2008PRB} In this subsection, we discuss the character and the origin of the nonthermal electron distribution during the photoinduced insulator-metal transition and the reopening of the gap. The high photon energy case shown in Figs.~\ref{fig:distrib} and~\ref{fig:set} is analysed as a specific example.

\begin{figure}
    \centering
    \includegraphics[width=1.0\hsize,clip]{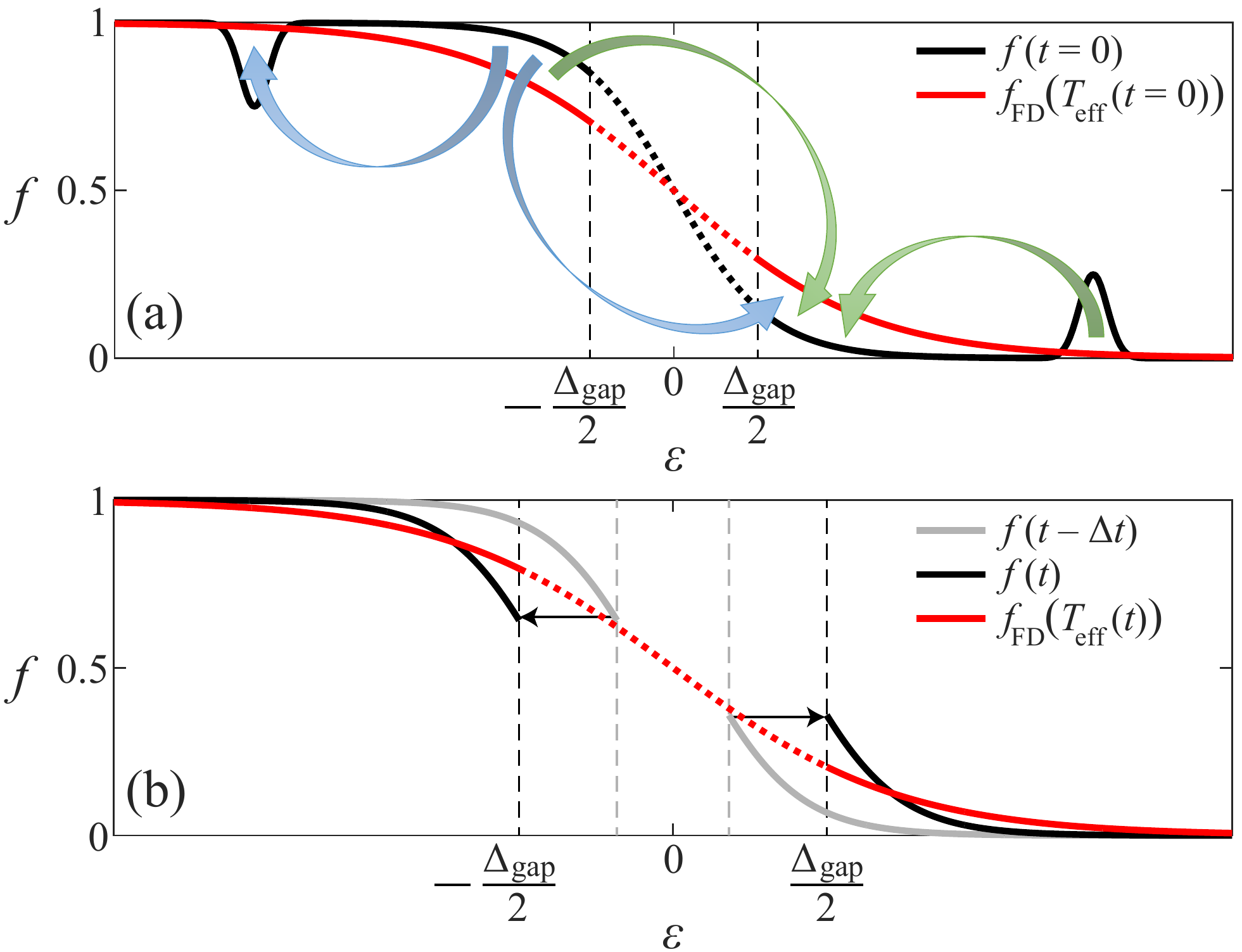}
    \caption{
    (a) Schematic diagram showing electron transfer from the lower to the upper band through two typical electron-electron scattering processes (blue and green arrows) right after the optical pump with $E_{\rm photon} \gg \Delta_{\rm gap}$. The black and the red lines represent the actual electron distribution and the corresponding Fermi-Dirac distribution, respectively. (b) Schematic diagram showing the effects of the gap reopening on $f_{\varepsilon}$. The gray line shows the actual distribution one time step $\Delta t$ earlier. The black and the red lines represent the actual and the Fermi-Dirac distributions.
    }\label{fig:sk}
\end{figure}

The distribution function $f(\varepsilon ,t=0)$ right after the optical pump and the equilibrium Fermi-Dirac distribution function $f_{\rm FD}(\varepsilon ,T_{\rm eff}(t=0))$ with the same energy are schematically drawn in Fig.~\ref{fig:sk}(a). Two typical electron-electron scattering processes that would change $f(\varepsilon ,t=0)$ closer to $f_{\rm FD}(\varepsilon, T_{\rm eff}(t=0))$ are also shown in green and blue arrows in Fig.~\ref{fig:sk}(a), which indicates the net electron transfer from the lower to the upper band responsible for the rapid rise of $\Delta N_{\rm upper}$ during the first 1~ps [Fig.~\ref{fig:set}(e)].

The reopening of the CO gap results in another stage with nonthermal electron distribution [Fig.~\ref{fig:set}(d)]. As shown in Fig.~\ref{fig:distrib}(b) for $t=18$~ps, the difference between the actual distribution and the thermal distribution is largest near the gap. Figure~\ref{fig:sk}(b) explains schematically why this happens. The distortion $u$ affects energy levels near the gap most sensitively. Adiabatic opening of the gap pushes electron and hole energy levels up without changing the occupation $f_{l \bf k}$ of the state $|l\bf k\rangle$, as indicated by arrows in Fig.~\ref{fig:sk}(b). This shift contributes to the increase of the effective temperature $T_{\rm eff}$ and competes against the cooling by the phonon thermal reservoir, which gives rise to a particularly slow electron energy relaxation of exponential decay rate $r\approx 0.013~{\rm ps^{-1}}$. [Fig.~\ref{fig:set}(a)].

The dynamics of the CO gap plays an essential role for the relaxation of the electronic energy shown in Fig.~\ref{fig:set}(a) in two aspects. First, the size of the CO gap directly affects the energy transfer from the electron to the phonon system, because phonons with energy smaller than the CO gap cannot participate in electron energy decay across the gap, limiting the thermal conductivity between the electron and the phonon system. This explains the order-of-magnitude increase of the relaxation time at the onset of the CO gap reopening. Second, the reopening of the CO gap pushes up the energy of the excited electrons and holes adiabatically, competing against the cooling of the electron system by the phonon thermal reservoir. The rapid increase of the CO gap makes the energy relaxation particularly slow right after $T_{\rm eff}$ drops below $T_{c}$.

\section{Comparison with Experiments}\label{sec:comparison}

We make comparisons between our results and experimental data. The photon energy of 1.55~eV used to melt CO in Ref.~\onlinecite{Beaud2014NATMAT} corresponds to the CO gap energy in $\rm Pr_{0.5}Ca_{0.5}MnO_{3}$, and, therefore, the experimental results can be compared with our results for $E_{\rm photon}=\Delta _{\rm gap}$. Approximately, the normalized off-resonance structural superlattice peak intensity and the normalized on-resonance charge order peak intensity in Fig.~2 in Ref.~\onlinecite{Beaud2014NATMAT} correspond to $\bar{u}^{2}$ and $\delta \bar{n}^{2}$ in our model. Large oscillation of the structural superlattice peak intensity and almost complete suppression of the CO peak intensity in the experiments are consistent with the evolution of $\bar{u}^{2}$ and $\delta \bar{n}^{2}$ shown in Fig.~\ref{fig:fluence}(e) for $E_{\rm photon}=\Delta_{\rm gap}$ case for our model.

Experimentally, unlike our simulation results, the energy for the photoinduced transition is not necessarily smaller than the energy for the thermodynamic transition.~\cite{Beaud2014NATMAT,Han2015SCIADV,Tao2016SCIREP} It has been proposed that the observed high critical fluence is related to the long wavelength distortions, or the changes in unit cell symmetry present in these materials, which cannot fully relax during the short time scale of photoinduced phase transitions. This discrepancy between simulation results and experimental results indicates that the long wavelength distortions, which are not included in the model, may indeed play an important role in the increase of the critical fluence, competing against the opposite effect from transient decoupling of incoherent phonons.

Increased energy efficiency of the photoinduced phase transition with a lower $E_{\rm photon}$ found from the simulations (Fig.~\ref{fig:thermo}) has been also observed in experiments. For example, in the CDW phases of $1T$-${\rm TaS_{2}}$ (Ref.~\onlinecite{Han2015SCIADV}) and $\rm VO_{2}$ (Ref.~\onlinecite{Tao2016SCIREP}), as $E_{\rm photon}$ is reduced from 1.5~eV to 0.5~eV, the energy required for the photoinduced transition drops by about 75\% and 50\% respectively, which are comparable to about 60\% drop of the critical fluence between $E_{\rm photon}\gg \Delta_{\rm gap}$ and $E_{\rm photon}= \Delta_{\rm gap}$ cases in the simulations.

\section{Summary}\label{sec:summary}

In summary, we have simulated photoinduced melting of charge order using a model of $M$$\rm O_{2}$ square lattice and a phonon thermal reservoir. The stages of CO melting, CO gap reopening, and thermal relaxation have been identified. During the stage of CO melting, the dynamics of the periodic lattice distortion is partially decoupled from and lags behind the dynamics of electronic order parameter due to large inertia of ions. As the effective electron temperature $T_{\rm eff}$ drops below $T_{c}$, electron system enters the stage of CO gap reopening and its state changes from thermal to nonthermal. The cooling rate of the electron system is reduced by an order of magnitude, because of adiabatic push of electron and hole levels by the gap reopening and reduced thermal conductivity between electron system and phonon thermal reservoir. During the stage of thermal relaxation, the electron system and the periodic lattice distortion maintain internal equilibrium, as they relax back to the initial state. Energy efficiency of photoinduced switching from insulator to metal more than doubles as the photon energy is reduced towards the CO gap. The frequency of coherent oscillation depends sensitively on time, fluence, and photon energy, which is interpreted in terms of the energy landscape dynamics. Agreements are found between the simulation results and recent experiments.

The results have shown intricately coupled dynamics of electrons, periodic lattice distortions, and incoherent phonons in nonequilibrium states excited by the optical pump in CO materials. Our approach can be extended in various ways, in particular by including more basis states in the tight-binding Hamiltonian. Inclusion of multiple orbitals would allow the study of the dynamics of orbital ordering in addition to the dynamics of charge ordering, as found in Ref.~\onlinecite{Beaud2014NATMAT}. Adding spin degrees of freedom and on-site Coulomb interaction~\cite{Moritz2013PRL} would allow the simulations of dynamics of magnetic ordering found in some CO or CDW materials. The study of such extended models would shed insight on how to make ultrafast switching devices out of CO or CDW materials.

\section*{Supplementary Material}\label{sec:supplementarymaterial}

See supplementary material for video simulations of the dynamics of electron distribution function $f(\varepsilon)$ together with the energy landscape $U(u)$ for $E_{\rm photon} \gg~\Delta_{\rm gap}$ and $E_{\rm photon} = \Delta_{\rm gap}$ cases.

\begin{acknowledgments}
L.Z. and K.H.A. were supported by the Computational Materials and Chemical Science Network under Grants No. DE-SC0007091 and 2014 NJIT Faculty Seed Grant. K.H.A. was further supported by 2013 Argonne X-ray Science Division Visitor Program. T.F.S. and M.v.V. are supported by the U.S. DOE, Office of Basic Energy Sciences, under Award No. DE-FG02-03ER46097. Work at Argonne National Laboratory was supported by the U.S. DOE, Office of Science, Office of Basic Energy Sciences, under Contract No. DE-AC02-06CH11357. The simulations primarily used computational resources managed by NJIT Academic and Research Computing Systems. This research also used resources of the National Energy Research Scientific Computing Center, supported by the U.S. DOE under Contract No. DE-AC02-05CH11231.
\end{acknowledgments}

\bibliography{DynCO-Ref}

\begin{thebibliography}{30}%
\makeatletter
\providecommand \@ifxundefined [1]{%
 \@ifx{#1\undefined}
}%
\providecommand \@ifnum [1]{%
 \ifnum #1\expandafter \@firstoftwo
 \else \expandafter \@secondoftwo
 \fi
}%
\providecommand \@ifx [1]{%
 \ifx #1\expandafter \@firstoftwo
 \else \expandafter \@secondoftwo
 \fi
}%
\providecommand \natexlab [1]{#1}%
\providecommand \enquote  [1]{``#1''}%
\providecommand \bibnamefont  [1]{#1}%
\providecommand \bibfnamefont [1]{#1}%
\providecommand \citenamefont [1]{#1}%
\providecommand \href@noop [0]{\@secondoftwo}%
\providecommand \href [0]{\begingroup \@sanitize@url \@href}%
\providecommand \@href[1]{\@@startlink{#1}\@@href}%
\providecommand \@@href[1]{\endgroup#1\@@endlink}%
\providecommand \@sanitize@url [0]{\catcode `\\12\catcode `\$12\catcode
  `\&12\catcode `\#12\catcode `\^12\catcode `\_12\catcode `\%12\relax}%
\providecommand \@@startlink[1]{}%
\providecommand \@@endlink[0]{}%
\providecommand \url  [0]{\begingroup\@sanitize@url \@url }%
\providecommand \@url [1]{\endgroup\@href {#1}{\urlprefix }}%
\providecommand \urlprefix  [0]{URL }%
\providecommand \Eprint [0]{\href }%
\providecommand \doibase [0]{http://dx.doi.org/}%
\providecommand \selectlanguage [0]{\@gobble}%
\providecommand \bibinfo  [0]{\@secondoftwo}%
\providecommand \bibfield  [0]{\@secondoftwo}%
\providecommand \translation [1]{[#1]}%
\providecommand \BibitemOpen [0]{}%
\providecommand \bibitemStop [0]{}%
\providecommand \bibitemNoStop [0]{.\EOS\space}%
\providecommand \EOS [0]{\spacefactor3000\relax}%
\providecommand \BibitemShut  [1]{\csname bibitem#1\endcsname}%
\let\auto@bib@innerbib\@empty
\bibitem [{\citenamefont {Rohwer}\ \emph {et~al.}(2011)\citenamefont {Rohwer},
  \citenamefont {Hellmann}, \citenamefont {Wiesenmayer}, \citenamefont {Sohrt},
  \citenamefont {Stange}, \citenamefont {Slomski}, \citenamefont {Carr},
  \citenamefont {Liu}, \citenamefont {Avila}, \citenamefont {Kallane},
  \citenamefont {Mathias}, \citenamefont {Kipp}, \citenamefont {Rossnagel},\
  and\ \citenamefont {Bauer}}]{Rohwer2011NATLET}%
  \BibitemOpen
  \bibfield  {author} {\bibinfo {author} {\bibfnamefont {T.}~\bibnamefont
  {Rohwer}}, \bibinfo {author} {\bibfnamefont {S.}~\bibnamefont {Hellmann}},
  \bibinfo {author} {\bibfnamefont {M.}~\bibnamefont {Wiesenmayer}}, \bibinfo
  {author} {\bibfnamefont {C.}~\bibnamefont {Sohrt}}, \bibinfo {author}
  {\bibfnamefont {A.}~\bibnamefont {Stange}}, \bibinfo {author} {\bibfnamefont
  {B.}~\bibnamefont {Slomski}}, \bibinfo {author} {\bibfnamefont
  {A.}~\bibnamefont {Carr}}, \bibinfo {author} {\bibfnamefont {Y.}~\bibnamefont
  {Liu}}, \bibinfo {author} {\bibfnamefont {L.~M.}\ \bibnamefont {Avila}},
  \bibinfo {author} {\bibfnamefont {M.}~\bibnamefont {Kallane}}, \bibinfo
  {author} {\bibfnamefont {S.}~\bibnamefont {Mathias}}, \bibinfo {author}
  {\bibfnamefont {L.}~\bibnamefont {Kipp}}, \bibinfo {author} {\bibfnamefont
  {K.}~\bibnamefont {Rossnagel}}, \ and\ \bibinfo {author} {\bibfnamefont
  {M.}~\bibnamefont {Bauer}},\ }\href {\doibase 10.1038/nature09829} {\bibfield
   {journal} {\bibinfo  {journal} {Nature (London)}\ }\textbf {\bibinfo
  {volume} {471}},\ \bibinfo {pages} {490} (\bibinfo {year}
  {2011})}\BibitemShut {NoStop}%
\bibitem [{\citenamefont {Cilento}\ \emph {et~al.}(2010)\citenamefont
  {Cilento}, \citenamefont {Giannetti}, \citenamefont {Ferrini}, \citenamefont
  {Conte}, \citenamefont {Sala}, \citenamefont {Coslovich}, \citenamefont
  {Rini}, \citenamefont {Cavalleri},\ and\ \citenamefont
  {Parmigiani}}]{Cilento2010JAP}%
  \BibitemOpen
  \bibfield  {author} {\bibinfo {author} {\bibfnamefont {F.}~\bibnamefont
  {Cilento}}, \bibinfo {author} {\bibfnamefont {C.}~\bibnamefont {Giannetti}},
  \bibinfo {author} {\bibfnamefont {G.}~\bibnamefont {Ferrini}}, \bibinfo
  {author} {\bibfnamefont {S.~D.}\ \bibnamefont {Conte}}, \bibinfo {author}
  {\bibfnamefont {T.}~\bibnamefont {Sala}}, \bibinfo {author} {\bibfnamefont
  {G.}~\bibnamefont {Coslovich}}, \bibinfo {author} {\bibfnamefont
  {M.}~\bibnamefont {Rini}}, \bibinfo {author} {\bibfnamefont {A.}~\bibnamefont
  {Cavalleri}}, \ and\ \bibinfo {author} {\bibfnamefont {F.}~\bibnamefont
  {Parmigiani}},\ }\href {\doibase 10.1063/1.3291105} {\bibfield  {journal}
  {\bibinfo  {journal} {Appl. Phys. Lett.}\ }\textbf {\bibinfo {volume} {96}},\
  \bibinfo {pages} {021102} (\bibinfo {year} {2010})}\BibitemShut {NoStop}%
\bibitem [{\citenamefont {Xue}\ \emph {et~al.}(2013)\citenamefont {Xue},
  \citenamefont {Jiang}, \citenamefont {Li}, \citenamefont {Lin}, \citenamefont
  {Ma},\ and\ \citenamefont {Jin}}]{Xue2013JAP}%
  \BibitemOpen
  \bibfield  {author} {\bibinfo {author} {\bibfnamefont {X.}~\bibnamefont
  {Xue}}, \bibinfo {author} {\bibfnamefont {M.}~\bibnamefont {Jiang}}, \bibinfo
  {author} {\bibfnamefont {G.}~\bibnamefont {Li}}, \bibinfo {author}
  {\bibfnamefont {X.}~\bibnamefont {Lin}}, \bibinfo {author} {\bibfnamefont
  {G.}~\bibnamefont {Ma}}, \ and\ \bibinfo {author} {\bibfnamefont
  {P.}~\bibnamefont {Jin}},\ }\href {\doibase 10.1063/1.4832776} {\bibfield
  {journal} {\bibinfo  {journal} {J. Appl. Phys.}\ }\textbf {\bibinfo {volume}
  {114}},\ \bibinfo {pages} {193506} (\bibinfo {year} {2013})}\BibitemShut
  {NoStop}%
\bibitem [{\citenamefont {Zhang}\ and\ \citenamefont
  {Averitt}(2014)}]{Zhang14REV}%
  \BibitemOpen
  \bibfield  {author} {\bibinfo {author} {\bibfnamefont {J.}~\bibnamefont
  {Zhang}}\ and\ \bibinfo {author} {\bibfnamefont {R.}~\bibnamefont
  {Averitt}},\ }\href
  {http://www.annualreviews.org/doi/abs/10.1146/annurev-matsci-070813-113258}
  {\bibfield  {journal} {\bibinfo  {journal} {Annu. Rev. Mater. Res.}\ }\textbf
  {\bibinfo {volume} {44}},\ \bibinfo {pages} {19} (\bibinfo {year}
  {2014})}\BibitemShut {NoStop}%
\bibitem [{\citenamefont {Kennes}\ \emph {et~al.}(2017)\citenamefont {Kennes},
  \citenamefont {Wilner}, \citenamefont {Reichman},\ and\ \citenamefont
  {Millis}}]{Kennes2017NAT}%
  \BibitemOpen
  \bibfield  {author} {\bibinfo {author} {\bibfnamefont {D.~M.}\ \bibnamefont
  {Kennes}}, \bibinfo {author} {\bibfnamefont {E.~Y.}\ \bibnamefont {Wilner}},
  \bibinfo {author} {\bibfnamefont {D.~R.}\ \bibnamefont {Reichman}}, \ and\
  \bibinfo {author} {\bibfnamefont {A.~J.}\ \bibnamefont {Millis}},\ }\href
  {http://dx.doi.org/10.1038/nphys4024} {\bibfield  {journal} {\bibinfo
  {journal} {Nat. Phys.}\ }\textbf {\bibinfo {volume} {13}},\ \bibinfo {pages}
  {479} (\bibinfo {year} {2017})}\BibitemShut {NoStop}%
\bibitem [{\citenamefont {Matsubara}\ \emph {et~al.}(2008)\citenamefont
  {Matsubara}, \citenamefont {Okimoto}, \citenamefont {Ogasawara},
  \citenamefont {Tomioka}, \citenamefont {Okamoto},\ and\ \citenamefont
  {Tokura}}]{Matsubara2008JAP}%
  \BibitemOpen
  \bibfield  {author} {\bibinfo {author} {\bibfnamefont {M.}~\bibnamefont
  {Matsubara}}, \bibinfo {author} {\bibfnamefont {Y.}~\bibnamefont {Okimoto}},
  \bibinfo {author} {\bibfnamefont {T.}~\bibnamefont {Ogasawara}}, \bibinfo
  {author} {\bibfnamefont {Y.}~\bibnamefont {Tomioka}}, \bibinfo {author}
  {\bibfnamefont {H.}~\bibnamefont {Okamoto}}, \ and\ \bibinfo {author}
  {\bibfnamefont {Y.}~\bibnamefont {Tokura}},\ }\href {\doibase
  10.1063/1.2836341} {\bibfield  {journal} {\bibinfo  {journal} {J. Appl.
  Phys.}\ }\textbf {\bibinfo {volume} {103}},\ \bibinfo {pages} {07B110}
  (\bibinfo {year} {2008})}\BibitemShut {NoStop}%
\bibitem [{\citenamefont {Piazza}\ \emph {et~al.}(2014)\citenamefont {Piazza},
  \citenamefont {Ma}, \citenamefont {Yang}, \citenamefont {Mann}, \citenamefont
  {Zhu}, \citenamefont {Li},\ and\ \citenamefont {Carbone}}]{Piazza2014STDYN}%
  \BibitemOpen
  \bibfield  {author} {\bibinfo {author} {\bibfnamefont {L.}~\bibnamefont
  {Piazza}}, \bibinfo {author} {\bibfnamefont {C.}~\bibnamefont {Ma}}, \bibinfo
  {author} {\bibfnamefont {H.~X.}\ \bibnamefont {Yang}}, \bibinfo {author}
  {\bibfnamefont {A.}~\bibnamefont {Mann}}, \bibinfo {author} {\bibfnamefont
  {Y.}~\bibnamefont {Zhu}}, \bibinfo {author} {\bibfnamefont {J.~Q.}\
  \bibnamefont {Li}}, \ and\ \bibinfo {author} {\bibfnamefont {F.}~\bibnamefont
  {Carbone}},\ }\href {https://www.ncbi.nlm.nih.gov/pmc/articles/PMC4711593/}
  {\bibfield  {journal} {\bibinfo  {journal} {Struct. Dyn.}\ }\textbf {\bibinfo
  {volume} {1}},\ \bibinfo {pages} {014501} (\bibinfo {year}
  {2014})}\BibitemShut {NoStop}%
\bibitem [{\citenamefont {Mankowsky}\ \emph {et~al.}(2015)\citenamefont
  {Mankowsky}, \citenamefont {F\"orst}, \citenamefont {Loew}, \citenamefont
  {Porras}, \citenamefont {Keimer},\ and\ \citenamefont
  {Cavalleri}}]{Mankowsky2015PRB}%
  \BibitemOpen
  \bibfield  {author} {\bibinfo {author} {\bibfnamefont {R.}~\bibnamefont
  {Mankowsky}}, \bibinfo {author} {\bibfnamefont {M.}~\bibnamefont {F\"orst}},
  \bibinfo {author} {\bibfnamefont {T.}~\bibnamefont {Loew}}, \bibinfo {author}
  {\bibfnamefont {J.}~\bibnamefont {Porras}}, \bibinfo {author} {\bibfnamefont
  {B.}~\bibnamefont {Keimer}}, \ and\ \bibinfo {author} {\bibfnamefont
  {A.}~\bibnamefont {Cavalleri}},\ }\href {\doibase 10.1103/PhysRevB.91.094308}
  {\bibfield  {journal} {\bibinfo  {journal} {Phys. Rev. B}\ }\textbf {\bibinfo
  {volume} {91}},\ \bibinfo {pages} {094308} (\bibinfo {year}
  {2015})}\BibitemShut {NoStop}%
\bibitem [{\citenamefont {Caviglia}\ \emph {et~al.}(2013)\citenamefont
  {Caviglia}, \citenamefont {F\"orst}, \citenamefont {Scherwitzl},
  \citenamefont {Khanna}, \citenamefont {Bromberger}, \citenamefont
  {Mankowsky}, \citenamefont {Singla}, \citenamefont {Chuang}, \citenamefont
  {Lee}, \citenamefont {Krupin}, \citenamefont {Schlotter}, \citenamefont
  {Turner}, \citenamefont {Dakovski}, \citenamefont {Minitti}, \citenamefont
  {Robinson}, \citenamefont {Scagnoli}, \citenamefont {Wilkins}, \citenamefont
  {Cavill}, \citenamefont {Gibert}, \citenamefont {Gariglio}, \citenamefont
  {Zubko}, \citenamefont {Triscone}, \citenamefont {Hill}, \citenamefont
  {Dhesi},\ and\ \citenamefont {Cavalleri}}]{Caviglia2013PRB}%
  \BibitemOpen
  \bibfield  {author} {\bibinfo {author} {\bibfnamefont {A.~D.}\ \bibnamefont
  {Caviglia}}, \bibinfo {author} {\bibfnamefont {M.}~\bibnamefont {F\"orst}},
  \bibinfo {author} {\bibfnamefont {R.}~\bibnamefont {Scherwitzl}}, \bibinfo
  {author} {\bibfnamefont {V.}~\bibnamefont {Khanna}}, \bibinfo {author}
  {\bibfnamefont {H.}~\bibnamefont {Bromberger}}, \bibinfo {author}
  {\bibfnamefont {R.}~\bibnamefont {Mankowsky}}, \bibinfo {author}
  {\bibfnamefont {R.}~\bibnamefont {Singla}}, \bibinfo {author} {\bibfnamefont
  {Y.-D.}\ \bibnamefont {Chuang}}, \bibinfo {author} {\bibfnamefont {W.~S.}\
  \bibnamefont {Lee}}, \bibinfo {author} {\bibfnamefont {O.}~\bibnamefont
  {Krupin}}, \bibinfo {author} {\bibfnamefont {W.~F.}\ \bibnamefont
  {Schlotter}}, \bibinfo {author} {\bibfnamefont {J.~J.}\ \bibnamefont
  {Turner}}, \bibinfo {author} {\bibfnamefont {G.~L.}\ \bibnamefont
  {Dakovski}}, \bibinfo {author} {\bibfnamefont {M.~P.}\ \bibnamefont
  {Minitti}}, \bibinfo {author} {\bibfnamefont {J.}~\bibnamefont {Robinson}},
  \bibinfo {author} {\bibfnamefont {V.}~\bibnamefont {Scagnoli}}, \bibinfo
  {author} {\bibfnamefont {S.~B.}\ \bibnamefont {Wilkins}}, \bibinfo {author}
  {\bibfnamefont {S.~A.}\ \bibnamefont {Cavill}}, \bibinfo {author}
  {\bibfnamefont {M.}~\bibnamefont {Gibert}}, \bibinfo {author} {\bibfnamefont
  {S.}~\bibnamefont {Gariglio}}, \bibinfo {author} {\bibfnamefont
  {P.}~\bibnamefont {Zubko}}, \bibinfo {author} {\bibfnamefont {J.-M.}\
  \bibnamefont {Triscone}}, \bibinfo {author} {\bibfnamefont {J.~P.}\
  \bibnamefont {Hill}}, \bibinfo {author} {\bibfnamefont {S.~S.}\ \bibnamefont
  {Dhesi}}, \ and\ \bibinfo {author} {\bibfnamefont {A.}~\bibnamefont
  {Cavalleri}},\ }\href {\doibase 10.1103/PhysRevB.88.220401} {\bibfield
  {journal} {\bibinfo  {journal} {Phys. Rev. B}\ }\textbf {\bibinfo {volume}
  {88}},\ \bibinfo {pages} {220401} (\bibinfo {year} {2013})}\BibitemShut
  {NoStop}%
\bibitem [{\citenamefont {Esposito}\ \emph {et~al.}(2017)\citenamefont
  {Esposito}, \citenamefont {Fechner}, \citenamefont {Mankowsky}, \citenamefont
  {Lemke}, \citenamefont {Chollet}, \citenamefont {Glownia}, \citenamefont
  {Nakamura}, \citenamefont {Kawasaki}, \citenamefont {Tokura}, \citenamefont
  {Staub}, \citenamefont {Beaud},\ and\ \citenamefont
  {F\"orst}}]{Esposito2017PRL}%
  \BibitemOpen
  \bibfield  {author} {\bibinfo {author} {\bibfnamefont {V.}~\bibnamefont
  {Esposito}}, \bibinfo {author} {\bibfnamefont {M.}~\bibnamefont {Fechner}},
  \bibinfo {author} {\bibfnamefont {R.}~\bibnamefont {Mankowsky}}, \bibinfo
  {author} {\bibfnamefont {H.}~\bibnamefont {Lemke}}, \bibinfo {author}
  {\bibfnamefont {M.}~\bibnamefont {Chollet}}, \bibinfo {author} {\bibfnamefont
  {J.~M.}\ \bibnamefont {Glownia}}, \bibinfo {author} {\bibfnamefont
  {M.}~\bibnamefont {Nakamura}}, \bibinfo {author} {\bibfnamefont
  {M.}~\bibnamefont {Kawasaki}}, \bibinfo {author} {\bibfnamefont
  {Y.}~\bibnamefont {Tokura}}, \bibinfo {author} {\bibfnamefont
  {U.}~\bibnamefont {Staub}}, \bibinfo {author} {\bibfnamefont
  {P.}~\bibnamefont {Beaud}}, \ and\ \bibinfo {author} {\bibfnamefont
  {M.}~\bibnamefont {F\"orst}},\ }\href {\doibase
  10.1103/PhysRevLett.118.247601} {\bibfield  {journal} {\bibinfo  {journal}
  {Phys. Rev. Lett.}\ }\textbf {\bibinfo {volume} {118}},\ \bibinfo {pages}
  {247601} (\bibinfo {year} {2017})}\BibitemShut {NoStop}%
\bibitem [{\citenamefont {Zhang}\ \emph {et~al.}(2013)\citenamefont {Zhang},
  \citenamefont {Li}, \citenamefont {Gray}, \citenamefont {He}, \citenamefont
  {Wang}, \citenamefont {Yang}, \citenamefont {Wang}, \citenamefont
  {Chakhalian},\ and\ \citenamefont {Xiao}}]{Zhang2013JAP}%
  \BibitemOpen
  \bibfield  {author} {\bibinfo {author} {\bibfnamefont {C.}~\bibnamefont
  {Zhang}}, \bibinfo {author} {\bibfnamefont {W.}~\bibnamefont {Li}}, \bibinfo
  {author} {\bibfnamefont {B.}~\bibnamefont {Gray}}, \bibinfo {author}
  {\bibfnamefont {B.}~\bibnamefont {He}}, \bibinfo {author} {\bibfnamefont
  {Y.}~\bibnamefont {Wang}}, \bibinfo {author} {\bibfnamefont {F.}~\bibnamefont
  {Yang}}, \bibinfo {author} {\bibfnamefont {X.}~\bibnamefont {Wang}}, \bibinfo
  {author} {\bibfnamefont {J.}~\bibnamefont {Chakhalian}}, \ and\ \bibinfo
  {author} {\bibfnamefont {M.}~\bibnamefont {Xiao}},\ }\href {\doibase
  10.1063/1.4793012} {\bibfield  {journal} {\bibinfo  {journal} {J. Appl.
  Phys.}\ }\textbf {\bibinfo {volume} {113}},\ \bibinfo {pages} {083901}
  (\bibinfo {year} {2013})}\BibitemShut {NoStop}%
\bibitem [{\citenamefont {Beaud}\ \emph {et~al.}(2014)\citenamefont {Beaud},
  \citenamefont {Caviezel}, \citenamefont {Mariager}, \citenamefont {Rettig},
  \citenamefont {Ingold}, \citenamefont {Dornes}, \citenamefont {Huang},
  \citenamefont {Johnson}, \citenamefont {Radovic}, \citenamefont {Huber},
  \citenamefont {Kubacka}, \citenamefont {Ferrer}, \citenamefont {Lemke},
  \citenamefont {Chollet}, \citenamefont {Zhu}, \citenamefont {Glownia},
  \citenamefont {Sikorski}, \citenamefont {Robert}, \citenamefont {Wadati},
  \citenamefont {Nakamura}, \citenamefont {Kawasaki}, \citenamefont {Tokura},
  \citenamefont {Johnson},\ and\ \citenamefont {Staub}}]{Beaud2014NATMAT}%
  \BibitemOpen
  \bibfield  {author} {\bibinfo {author} {\bibfnamefont {P.}~\bibnamefont
  {Beaud}}, \bibinfo {author} {\bibfnamefont {A.}~\bibnamefont {Caviezel}},
  \bibinfo {author} {\bibfnamefont {S.~O.}\ \bibnamefont {Mariager}}, \bibinfo
  {author} {\bibfnamefont {L.}~\bibnamefont {Rettig}}, \bibinfo {author}
  {\bibfnamefont {G.}~\bibnamefont {Ingold}}, \bibinfo {author} {\bibfnamefont
  {C.}~\bibnamefont {Dornes}}, \bibinfo {author} {\bibfnamefont {S.-W.}\
  \bibnamefont {Huang}}, \bibinfo {author} {\bibfnamefont {J.~A.}\ \bibnamefont
  {Johnson}}, \bibinfo {author} {\bibfnamefont {M.}~\bibnamefont {Radovic}},
  \bibinfo {author} {\bibfnamefont {T.}~\bibnamefont {Huber}}, \bibinfo
  {author} {\bibfnamefont {T.}~\bibnamefont {Kubacka}}, \bibinfo {author}
  {\bibfnamefont {A.}~\bibnamefont {Ferrer}}, \bibinfo {author} {\bibfnamefont
  {H.~T.}\ \bibnamefont {Lemke}}, \bibinfo {author} {\bibfnamefont
  {M.}~\bibnamefont {Chollet}}, \bibinfo {author} {\bibfnamefont
  {D.}~\bibnamefont {Zhu}}, \bibinfo {author} {\bibfnamefont {J.~M.}\
  \bibnamefont {Glownia}}, \bibinfo {author} {\bibfnamefont {M.}~\bibnamefont
  {Sikorski}}, \bibinfo {author} {\bibfnamefont {A.}~\bibnamefont {Robert}},
  \bibinfo {author} {\bibfnamefont {H.}~\bibnamefont {Wadati}}, \bibinfo
  {author} {\bibfnamefont {M.}~\bibnamefont {Nakamura}}, \bibinfo {author}
  {\bibfnamefont {M.}~\bibnamefont {Kawasaki}}, \bibinfo {author}
  {\bibfnamefont {Y.}~\bibnamefont {Tokura}}, \bibinfo {author} {\bibfnamefont
  {S.~L.}\ \bibnamefont {Johnson}}, \ and\ \bibinfo {author} {\bibfnamefont
  {U.}~\bibnamefont {Staub}},\ }\href {http://dx.doi.org/10.1038/nmat4046}
  {\bibfield  {journal} {\bibinfo  {journal} {Nat. Mater.}\ }\textbf {\bibinfo
  {volume} {13}},\ \bibinfo {pages} {923} (\bibinfo {year} {2014})}\BibitemShut
  {NoStop}%
\bibitem [{\citenamefont {Shao}\ \emph {et~al.}(2016)\citenamefont {Shao},
  \citenamefont {Xiao}, \citenamefont {Lu}, \citenamefont {Lv}, \citenamefont
  {Li}, \citenamefont {Zhu},\ and\ \citenamefont {Sun}}]{Shao2016PRB}%
  \BibitemOpen
  \bibfield  {author} {\bibinfo {author} {\bibfnamefont {D.~F.}\ \bibnamefont
  {Shao}}, \bibinfo {author} {\bibfnamefont {R.~C.}\ \bibnamefont {Xiao}},
  \bibinfo {author} {\bibfnamefont {W.~J.}\ \bibnamefont {Lu}}, \bibinfo
  {author} {\bibfnamefont {H.~Y.}\ \bibnamefont {Lv}}, \bibinfo {author}
  {\bibfnamefont {J.~Y.}\ \bibnamefont {Li}}, \bibinfo {author} {\bibfnamefont
  {X.~B.}\ \bibnamefont {Zhu}}, \ and\ \bibinfo {author} {\bibfnamefont
  {Y.~P.}\ \bibnamefont {Sun}},\ }\href {\doibase 10.1103/PhysRevB.94.125126}
  {\bibfield  {journal} {\bibinfo  {journal} {Phys. Rev. B}\ }\textbf {\bibinfo
  {volume} {94}},\ \bibinfo {pages} {125126} (\bibinfo {year}
  {2016})}\BibitemShut {NoStop}%
\bibitem [{\citenamefont {van Veenendaal}(2013)}]{vanVeenendaal2013PRB}%
  \BibitemOpen
  \bibfield  {author} {\bibinfo {author} {\bibfnamefont {M.}~\bibnamefont {van
  Veenendaal}},\ }\href {\doibase 10.1103/PhysRevB.87.235118} {\bibfield
  {journal} {\bibinfo  {journal} {Phys. Rev. B}\ }\textbf {\bibinfo {volume}
  {87}},\ \bibinfo {pages} {235118} (\bibinfo {year} {2013})}\BibitemShut
  {NoStop}%
\bibitem [{\citenamefont {Groeneveld}, \citenamefont {Sprik},\ and\
  \citenamefont {Lagendijk}(1992)}]{Groeneveld1992PRB}%
  \BibitemOpen
  \bibfield  {author} {\bibinfo {author} {\bibfnamefont {R.~H.~M.}\
  \bibnamefont {Groeneveld}}, \bibinfo {author} {\bibfnamefont
  {R.}~\bibnamefont {Sprik}}, \ and\ \bibinfo {author} {\bibfnamefont
  {A.}~\bibnamefont {Lagendijk}},\ }\href {\doibase 10.1103/PhysRevB.45.5079}
  {\bibfield  {journal} {\bibinfo  {journal} {Phys. Rev. B}\ }\textbf {\bibinfo
  {volume} {45}},\ \bibinfo {pages} {5079} (\bibinfo {year}
  {1992})}\BibitemShut {NoStop}%
\bibitem [{\citenamefont {Groeneveld}, \citenamefont {Sprik},\ and\
  \citenamefont {Lagendijk}(1995)}]{Groeneveld1995PRB}%
  \BibitemOpen
  \bibfield  {author} {\bibinfo {author} {\bibfnamefont {R.~H.~M.}\
  \bibnamefont {Groeneveld}}, \bibinfo {author} {\bibfnamefont
  {R.}~\bibnamefont {Sprik}}, \ and\ \bibinfo {author} {\bibfnamefont
  {A.}~\bibnamefont {Lagendijk}},\ }\href {\doibase 10.1103/PhysRevB.51.11433}
  {\bibfield  {journal} {\bibinfo  {journal} {Phys. Rev. B}\ }\textbf {\bibinfo
  {volume} {51}},\ \bibinfo {pages} {11433} (\bibinfo {year}
  {1995})}\BibitemShut {NoStop}%
\bibitem [{\citenamefont {Rethfeld}\ \emph {et~al.}(2002)\citenamefont
  {Rethfeld}, \citenamefont {Kaiser}, \citenamefont {Vicanek},\ and\
  \citenamefont {Simon}}]{Rethfeld2002PRB}%
  \BibitemOpen
  \bibfield  {author} {\bibinfo {author} {\bibfnamefont {B.}~\bibnamefont
  {Rethfeld}}, \bibinfo {author} {\bibfnamefont {A.}~\bibnamefont {Kaiser}},
  \bibinfo {author} {\bibfnamefont {M.}~\bibnamefont {Vicanek}}, \ and\
  \bibinfo {author} {\bibfnamefont {G.}~\bibnamefont {Simon}},\ }\href
  {\doibase 10.1103/PhysRevB.65.214303} {\bibfield  {journal} {\bibinfo
  {journal} {Phys. Rev. B}\ }\textbf {\bibinfo {volume} {65}},\ \bibinfo
  {pages} {214303} (\bibinfo {year} {2002})}\BibitemShut {NoStop}%
\bibitem [{\citenamefont {Demsar}\ \emph {et~al.}(2003)\citenamefont {Demsar},
  \citenamefont {Averitt}, \citenamefont {Ahn}, \citenamefont {Graf},
  \citenamefont {Trugman}, \citenamefont {Kabanov}, \citenamefont {Sarrao},\
  and\ \citenamefont {Taylor}}]{Demsar2003PRL}%
  \BibitemOpen
  \bibfield  {author} {\bibinfo {author} {\bibfnamefont {J.}~\bibnamefont
  {Demsar}}, \bibinfo {author} {\bibfnamefont {R.~D.}\ \bibnamefont {Averitt}},
  \bibinfo {author} {\bibfnamefont {K.~H.}\ \bibnamefont {Ahn}}, \bibinfo
  {author} {\bibfnamefont {M.~J.}\ \bibnamefont {Graf}}, \bibinfo {author}
  {\bibfnamefont {S.~A.}\ \bibnamefont {Trugman}}, \bibinfo {author}
  {\bibfnamefont {V.~V.}\ \bibnamefont {Kabanov}}, \bibinfo {author}
  {\bibfnamefont {J.~L.}\ \bibnamefont {Sarrao}}, \ and\ \bibinfo {author}
  {\bibfnamefont {A.~J.}\ \bibnamefont {Taylor}},\ }\href {\doibase
  10.1103/PhysRevLett.91.027401} {\bibfield  {journal} {\bibinfo  {journal}
  {Phys. Rev. Lett.}\ }\textbf {\bibinfo {volume} {91}},\ \bibinfo {pages}
  {027401} (\bibinfo {year} {2003})}\BibitemShut {NoStop}%
\bibitem [{\citenamefont {Ahn}\ \emph {et~al.}(2004)\citenamefont {Ahn},
  \citenamefont {Graf}, \citenamefont {Trugman}, \citenamefont {Demsar},
  \citenamefont {Averitt}, \citenamefont {Sarrao},\ and\ \citenamefont
  {Taylor}}]{Ahn2004PRB}%
  \BibitemOpen
  \bibfield  {author} {\bibinfo {author} {\bibfnamefont {K.~H.}\ \bibnamefont
  {Ahn}}, \bibinfo {author} {\bibfnamefont {M.~J.}\ \bibnamefont {Graf}},
  \bibinfo {author} {\bibfnamefont {S.~A.}\ \bibnamefont {Trugman}}, \bibinfo
  {author} {\bibfnamefont {J.}~\bibnamefont {Demsar}}, \bibinfo {author}
  {\bibfnamefont {R.~D.}\ \bibnamefont {Averitt}}, \bibinfo {author}
  {\bibfnamefont {J.~L.}\ \bibnamefont {Sarrao}}, \ and\ \bibinfo {author}
  {\bibfnamefont {A.~J.}\ \bibnamefont {Taylor}},\ }\href {\doibase
  10.1103/PhysRevB.69.045114} {\bibfield  {journal} {\bibinfo  {journal} {Phys.
  Rev. B}\ }\textbf {\bibinfo {volume} {69}},\ \bibinfo {pages} {045114}
  (\bibinfo {year} {2004})}\BibitemShut {NoStop}%
\bibitem [{\citenamefont {Ahn}\ and\ \citenamefont
  {Millis}(2000)}]{Ahn2000PRB}%
  \BibitemOpen
  \bibfield  {author} {\bibinfo {author} {\bibfnamefont {K.~H.}\ \bibnamefont
  {Ahn}}\ and\ \bibinfo {author} {\bibfnamefont {A.~J.}\ \bibnamefont
  {Millis}},\ }\href {\doibase 10.1103/PhysRevB.61.13545} {\bibfield  {journal}
  {\bibinfo  {journal} {Phys. Rev. B}\ }\textbf {\bibinfo {volume} {61}},\
  \bibinfo {pages} {13545} (\bibinfo {year} {2000})}\BibitemShut {NoStop}%
\bibitem [{\citenamefont {Ahn}\ and\ \citenamefont
  {Millis}(2001)}]{Ahn2001PRB}%
  \BibitemOpen
  \bibfield  {author} {\bibinfo {author} {\bibfnamefont {K.~H.}\ \bibnamefont
  {Ahn}}\ and\ \bibinfo {author} {\bibfnamefont {A.~J.}\ \bibnamefont
  {Millis}},\ }\href {\doibase 10.1103/PhysRevB.64.115103} {\bibfield
  {journal} {\bibinfo  {journal} {Phys. Rev. B}\ }\textbf {\bibinfo {volume}
  {64}},\ \bibinfo {pages} {115103} (\bibinfo {year} {2001})}\BibitemShut
  {NoStop}%
\bibitem [{\citenamefont {Demsar}, \citenamefont
  {Biljakovi\ifmmode~\acute{c}\else \'{c}\fi{}},\ and\ \citenamefont
  {Mihailovic}(1999)}]{Demsar1999PRL}%
  \BibitemOpen
  \bibfield  {author} {\bibinfo {author} {\bibfnamefont {J.}~\bibnamefont
  {Demsar}}, \bibinfo {author} {\bibfnamefont {K.}~\bibnamefont
  {Biljakovi\ifmmode~\acute{c}\else \'{c}\fi{}}}, \ and\ \bibinfo {author}
  {\bibfnamefont {D.}~\bibnamefont {Mihailovic}},\ }\href {\doibase
  10.1103/PhysRevLett.83.800} {\bibfield  {journal} {\bibinfo  {journal} {Phys.
  Rev. Lett.}\ }\textbf {\bibinfo {volume} {83}},\ \bibinfo {pages} {800}
  (\bibinfo {year} {1999})}\BibitemShut {NoStop}%
\bibitem [{\citenamefont {Han}\ \emph {et~al.}(2012)\citenamefont {Han},
  \citenamefont {Tao}, \citenamefont {Mahanti}, \citenamefont {Chang},
  \citenamefont {Ruan}, \citenamefont {Malliakas},\ and\ \citenamefont
  {Kanatzidis}}]{Han2012PRB}%
  \BibitemOpen
  \bibfield  {author} {\bibinfo {author} {\bibfnamefont {T.-R.~T.}\
  \bibnamefont {Han}}, \bibinfo {author} {\bibfnamefont {Z.}~\bibnamefont
  {Tao}}, \bibinfo {author} {\bibfnamefont {S.~D.}\ \bibnamefont {Mahanti}},
  \bibinfo {author} {\bibfnamefont {K.}~\bibnamefont {Chang}}, \bibinfo
  {author} {\bibfnamefont {C.-Y.}\ \bibnamefont {Ruan}}, \bibinfo {author}
  {\bibfnamefont {C.~D.}\ \bibnamefont {Malliakas}}, \ and\ \bibinfo {author}
  {\bibfnamefont {M.~G.}\ \bibnamefont {Kanatzidis}},\ }\href {\doibase
  10.1103/PhysRevB.86.075145} {\bibfield  {journal} {\bibinfo  {journal} {Phys.
  Rev. B}\ }\textbf {\bibinfo {volume} {86}},\ \bibinfo {pages} {075145}
  (\bibinfo {year} {2012})}\BibitemShut {NoStop}%
\bibitem [{\citenamefont {Gr\"uner}(1994)}]{GrunerBook}%
  \BibitemOpen
  \bibfield  {author} {\bibinfo {author} {\bibfnamefont {G.}~\bibnamefont
  {Gr\"uner}},\ }\href@noop {} {\emph {\bibinfo {title} {Density Waves in
  Solids}}}\ (\bibinfo  {publisher} {Addison-Wesley Publishing Company, Los
  Angeles},\ \bibinfo {year} {1994})\BibitemShut {NoStop}%
\bibitem [{\citenamefont {Woll}\ and\ \citenamefont {Kohn}(1962)}]{Woll1962PR}%
  \BibitemOpen
  \bibfield  {author} {\bibinfo {author} {\bibfnamefont {E.~J.}\ \bibnamefont
  {Woll}}\ and\ \bibinfo {author} {\bibfnamefont {W.}~\bibnamefont {Kohn}},\
  }\href {\doibase 10.1103/PhysRev.126.1693} {\bibfield  {journal} {\bibinfo
  {journal} {Phys. Rev.}\ }\textbf {\bibinfo {volume} {126}},\ \bibinfo {pages}
  {1693} (\bibinfo {year} {1962})}\BibitemShut {NoStop}%
\bibitem [{\citenamefont {Allen}(1987)}]{Allen1987PRL}%
  \BibitemOpen
  \bibfield  {author} {\bibinfo {author} {\bibfnamefont {P.~B.}\ \bibnamefont
  {Allen}},\ }\href {\doibase 10.1103/PhysRevLett.59.1460} {\bibfield
  {journal} {\bibinfo  {journal} {Phys. Rev. Lett.}\ }\textbf {\bibinfo
  {volume} {59}},\ \bibinfo {pages} {1460} (\bibinfo {year}
  {1987})}\BibitemShut {NoStop}%
\bibitem [{\citenamefont {Kabanov}\ and\ \citenamefont
  {Alexandrov}(2008)}]{Kabanov2008PRB}%
  \BibitemOpen
  \bibfield  {author} {\bibinfo {author} {\bibfnamefont {V.~V.}\ \bibnamefont
  {Kabanov}}\ and\ \bibinfo {author} {\bibfnamefont {A.~S.}\ \bibnamefont
  {Alexandrov}},\ }\href {\doibase 10.1103/PhysRevB.78.174514} {\bibfield
  {journal} {\bibinfo  {journal} {Phys. Rev. B}\ }\textbf {\bibinfo {volume}
  {78}},\ \bibinfo {pages} {174514} (\bibinfo {year} {2008})}\BibitemShut
  {NoStop}%
\bibitem [{\citenamefont {Han}\ \emph {et~al.}(2015)\citenamefont {Han},
  \citenamefont {Zhou}, \citenamefont {Malliakas}, \citenamefont {Duxbury},
  \citenamefont {Mahanti}, \citenamefont {Kanatzidis},\ and\ \citenamefont
  {Ruan}}]{Han2015SCIADV}%
  \BibitemOpen
  \bibfield  {author} {\bibinfo {author} {\bibfnamefont {T.-R.~T.}\
  \bibnamefont {Han}}, \bibinfo {author} {\bibfnamefont {F.}~\bibnamefont
  {Zhou}}, \bibinfo {author} {\bibfnamefont {C.~D.}\ \bibnamefont {Malliakas}},
  \bibinfo {author} {\bibfnamefont {P.~M.}\ \bibnamefont {Duxbury}}, \bibinfo
  {author} {\bibfnamefont {S.~D.}\ \bibnamefont {Mahanti}}, \bibinfo {author}
  {\bibfnamefont {M.~G.}\ \bibnamefont {Kanatzidis}}, \ and\ \bibinfo {author}
  {\bibfnamefont {C.-Y.}\ \bibnamefont {Ruan}},\ }\href {\doibase
  10.1126/sciadv.1400173} {\bibfield  {journal} {\bibinfo  {journal} {Sci.
  Adv.}\ }\textbf {\bibinfo {volume} {1}},\ \bibinfo {pages} {e1400173}
  (\bibinfo {year} {2015})}\BibitemShut {NoStop}%
\bibitem [{\citenamefont {Tao}\ \emph {et~al.}(2016)\citenamefont {Tao},
  \citenamefont {Zhou}, \citenamefont {Han}, \citenamefont {Torres},
  \citenamefont {Wang}, \citenamefont {Sepulveda}, \citenamefont {Chang},
  \citenamefont {Young}, \citenamefont {Lunt},\ and\ \citenamefont
  {Ruan}}]{Tao2016SCIREP}%
  \BibitemOpen
  \bibfield  {author} {\bibinfo {author} {\bibfnamefont {Z.}~\bibnamefont
  {Tao}}, \bibinfo {author} {\bibfnamefont {F.}~\bibnamefont {Zhou}}, \bibinfo
  {author} {\bibfnamefont {T.-R.~T.}\ \bibnamefont {Han}}, \bibinfo {author}
  {\bibfnamefont {D.}~\bibnamefont {Torres}}, \bibinfo {author} {\bibfnamefont
  {T.}~\bibnamefont {Wang}}, \bibinfo {author} {\bibfnamefont {N.}~\bibnamefont
  {Sepulveda}}, \bibinfo {author} {\bibfnamefont {K.}~\bibnamefont {Chang}},
  \bibinfo {author} {\bibfnamefont {M.}~\bibnamefont {Young}}, \bibinfo
  {author} {\bibfnamefont {R.~R.}\ \bibnamefont {Lunt}}, \ and\ \bibinfo
  {author} {\bibfnamefont {C.-Y.}\ \bibnamefont {Ruan}},\ }\href
  {http://dx.doi.org/10.1038/srep38514} {\bibfield  {journal} {\bibinfo
  {journal} {Sci. Rep.}\ }\textbf {\bibinfo {volume} {6}},\ \bibinfo {pages}
  {38514} (\bibinfo {year} {2016})}\BibitemShut {NoStop}%
\bibitem [{\citenamefont {Moritz}\ \emph {et~al.}(2013)\citenamefont {Moritz},
  \citenamefont {Kemper}, \citenamefont {Sentef}, \citenamefont {Devereaux},\
  and\ \citenamefont {Freericks}}]{Moritz2013PRL}%
  \BibitemOpen
  \bibfield  {author} {\bibinfo {author} {\bibfnamefont {B.}~\bibnamefont
  {Moritz}}, \bibinfo {author} {\bibfnamefont {A.~F.}\ \bibnamefont {Kemper}},
  \bibinfo {author} {\bibfnamefont {M.}~\bibnamefont {Sentef}}, \bibinfo
  {author} {\bibfnamefont {T.~P.}\ \bibnamefont {Devereaux}}, \ and\ \bibinfo
  {author} {\bibfnamefont {J.~K.}\ \bibnamefont {Freericks}},\ }\href {\doibase
  10.1103/PhysRevLett.111.077401} {\bibfield  {journal} {\bibinfo  {journal}
  {Phys. Rev. Lett.}\ }\textbf {\bibinfo {volume} {111}},\ \bibinfo {pages}
  {077401} (\bibinfo {year} {2013})}\BibitemShut {NoStop}%
\end{thebibliography}%

\end{document}